\documentclass[12pt]{article}
\usepackage{amsmath,amssymb,exscale}
\usepackage{graphicx}

\newcommand{\bra}[1]{\langle #1 |}
\newcommand{\ket}[1]{| #1 \rangle}
\newcommand{\braket}[2]{\langle #1 | #2 \rangle}

\newcommand{\Eq}[1]{Eq.~(\ref{#1})}
\newcommand{\Sec}[1]{Sec.~\ref{Sec:#1}}
\newcommand{\Fig}[1]{Fig.~\ref{Fig:#1}}
\newcommand{\op}[1]{\boldsymbol{#1}}
\newcommand\Hil{\mathcal H}

\newcommand{\xplus}{\mathbin{\mkern 2mu{+}\mkern -16mu{\bigcirc}}} 

\newcommand{\Mat}[1]{\left(\begin{array}{cccccccccc}#1\end{array}\right)} 
\newcommand{\mat}[1]{\left(\begin{smallmatrix}#1\end{smallmatrix}\right)} 
\newcommand{\Col}[1]{\begin{array}{l}#1\end{array}} 
\newcommand{\cNOT}[1]{\mbox{\sf c-NOT$\!_{\mathnormal{#1}}$}}
\newcommand{\nrm}[1]{\frac{\sqrt{#1}}{\,#1}}

\newcommand{\twox}[2]{\mbox{\vbox{\hbox{#1}\hbox{#2}}}}

\newcommand{\IorII}[2]{#1}

\newlength{\myparwidth}
\newlength{\myparindent}

\newcommand{\rightpic}[3]{{
\setlength{\myparwidth}{\textwidth}
\setlength{\myparindent}{\parindent}
\addtolength{\myparwidth}{-#1}
\addtolength{\myparwidth}{-3pt}
\smallskip
\noindent
\parbox{\myparwidth}{
\hspace*{\myparindent}#3}
\parbox{#1}{#2}
\smallskip
}}

\newenvironment{Note}{\small\smallskip \noindent\textbullet\ {\em Note:} }{\smallskip}

\def\paper{paper} 

\title{Programmable Quantum Networks with Pure States}

\author{{\em Alexander Yu.\ Vlasov
}
}

\begin{document}
\sloppy
\maketitle


\begin{abstract}
Modern classical computing devices, except of simplest
calculators, have von Neumann architecture, {\em i.e.}, a part of the memory
is used for the program and a part for the data. It is likely, that analogues 
of such architecture are also desirable for the future applications in 
quantum computing, communications and control. It is also interesting 
for the modern theoretical research in the quantum information science 
and raises challenging questions about an experimental assessment of 
such a programmable models. Together with some progress in the given 
direction, such ideas encounter specific problems arising from the 
very essence of quantum laws. Currently are known two different ways 
to overcome such problems, sometime denoted as a stochastic and 
deterministic approach. The presented {\paper} is devoted to the
second one, that is also may be called {\em the programmable quantum 
networks with pure states}.

In the {\paper} are discussed basic principles and theoretical models 
that can be used for the design of such nano-devices, {\em e.g.},
the conditional quantum dynamics, the Nielsen-Chuang ``no-programming
theorem,'' the idea of deterministic and stochastic quantum gates
arrays. Both programmable quantum networks with finite registers 
and hybrid models with continuous quantum variables are considered. 
As a basic model for the universal programmable quantum network with pure
states and finite program register is chosen a ``{\sc Control-Shift}''
quantum processor architecture with three buses introduced in earlier works. 
It is shown also, that quantum cellular automata approach to the
construction of an universal programmable quantum computer often 
may be considered as the particular case of such design. 
\end{abstract}

\section{Introduction}
\label{Sec:Intro}

There are two almost independent ways of the classification of 
the quantum computational networks. In this paper and
in many other works about the theory of quantum computations are used
models with quantum networks acting on pure states represented 
as vectors in Hilbert spaces \cite{QCN}. It is also possible
to consider mixed states and to use density matrices \cite{QCmix}.
Such methods are especially useful in theory of open quantum
systems, but for networks described in this {\paper} it is enough
to consider only pure states.

\begin{figure}[htb]
\begin{center}
\parbox{0.49\textwidth}{\centering
\IorII{\includegraphics[scale=0.8]{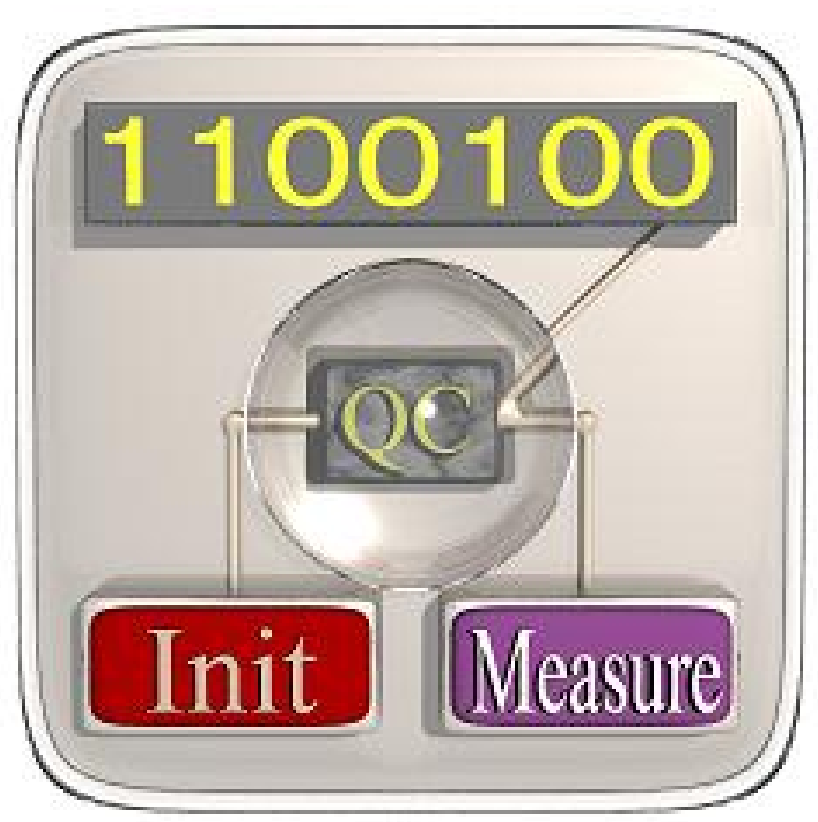}\\}
{\includegraphics[bb=0 0 480 480,scale=0.4]{qnet1b.jpg}\\}
a)}%
\parbox{0.49\textwidth}{\centering
\IorII{\includegraphics[scale=0.8]{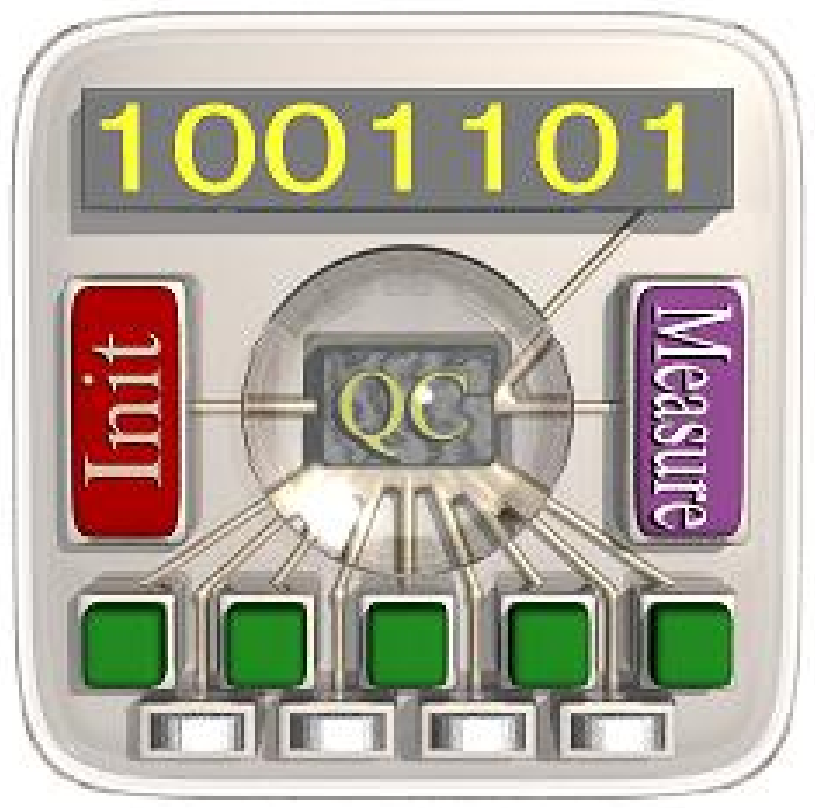}\\}
{\includegraphics[bb=0 0 480 480,scale=0.4]{qnet2b.jpg}\\}
b)}
\end{center}
\caption{Usual schemes of quantum networks.
a) The simplest quantum network with a fixed structure. 
b) The network with a set of gates for external control.}
\label{Fig:qnet12}
\end{figure}

On the other hand, the {\em structure} of a quantum network may be specified 
using few different levels, which also have some analogue with the classical case. 
The simplest case --- is a fixed network for the resolution of a particular 
task, {\em e.g.}, the quantum network for Shor's factoring algorithm 
\cite{Shor}. In the quantum case there are three different steps: the 
{\em preparation} of initial state, the quantum {\em evolution} described 
by given quantum network, and the {\em measurement} (\Fig{qnet12}a).

More difficult level --- is a network with tuneable and interchangeable elements
for managing with different tasks (\Fig{qnet12}b). It is common also for most 
experiments. Here is also present the initialization and the read-out (measurement), 
but between them may be considered different sequences of operations. 
Already in the earliest works \cite{UQC} was raised a question
about an universal set of such elementary operations, {\em the quantum gates}.

Even the second kind of network still rather resembles a quantum ``calculator'' 
\Fig{qnet12}b than a computer, because it is controlled via external 
manipulations instead of doing some program. It is especially essential
for the quantum case, because such a control here is described as a classical
process and so there is an additional difficulty for the unified description
of the whole system.

\begin{figure}[htb]
\begin{center}
\IorII{\includegraphics[scale=1.25]{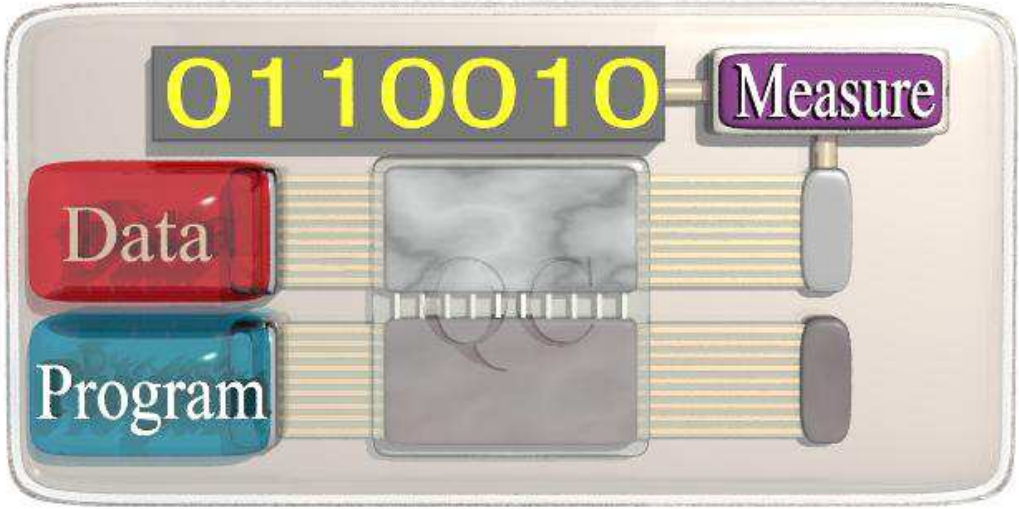}}
{\includegraphics[bb=0 0 800 400,scale=0.4]{qnet3b.jpg}}
\end{center}
\caption{The programmable quantum network.}
\label{Fig:qnet3}
\end{figure}

It is possible to consider a next level --- the programmable quantum network
with two subsystems: {\em the data} and {\em the program} \Fig{qnet3}. It is 
similar with von Neumann architecture of the classical computers \cite{EDVAC}.

With the usual notation of the quantum computation the programmable quantum
networks $\mathsf U$ (with pure states) may be described as:
\begin{equation}
\mathsf U\bigl(\ket{D}\ket{\varPi}\bigr) =
  \bigl(\op u_\varPi\ket{D}\bigr)\,\ket{\varPi}.
\label{qprog}
\end{equation}
Here $\ket{D}$ and $\ket{\varPi}$ are states of {\em data} and {\em program}
registers before operation. After the application of the {\em fixed} unitary
operator $\mathsf U$, the state of the data register may be described as 
$\op u_\varPi\ket{D}$,
{\em i.e.}, some operator $\op u_\varPi$ is applied to the data state and
it depends on the state $\ket{\varPi}$ of the program register. State of the program
register after the operation $\mathsf U$ is not changed (a more general case
is represented in \Eq{qprog'} below on the page \pageref{qprog'}).

It was found \cite{NC97} that for the pure states \Eq{qprog} may be valid only 
if all different states $\ket{\varPi}$ of the program register {\em are orthogonal}.
{\em E.g.}, if there is some program (state) $\ket{\varPi_1}$ for the implementation
of an operator $\op u_1$ and we need to implement another operator $\op u_2$
using some program $\ket{\varPi_2}$, then: 
\begin{equation}
 \op u_1 \ne \op u_2 \quad \Longrightarrow 
 \quad \braket{\varPi_1}{\varPi_2} = 0.
\label{ortprog}
\end{equation}
The \Eq{ortprog} was derived in \cite{NC97} from \Eq{qprog} and 
unitarity of $\mathsf U$ (see \Sec{nogo}).

It was considered in \cite{NC97} as some kind of {\em the no-go theorem} for 
universality of a programmable quantum network, because in such a case 
dimension of the Hilbert space for the program register must be equal to the number 
of different programs, {\em i.e.}, unitary operators, but for the {\em exactly
universal} quantum computer there are {\em infinite} number of such programs. 

There are a few important no-go results in the theory of quantum computing
and most known --- is a {\em no-cloning} theorem \cite{noclon}. 
The no-cloning theorem may be deduced from {\em linearity} of quantum
mechanics, and the just mentioned ``no-programming'' problem even more 
subtle, {\em e.g.}, exists a {\em linear non-unitary} operator satisfying
\Eq{qprog} with dimension of the program register only in two times bigger
than the data one \cite{QPC03} (see \Sec{non}).
The problem with infinite dimension of the program register in the universal
programmable quantum network appears due to the orthogonality of the different 
program states \Eq{ortprog} derived from unitarity of {\sf U}.

The no-programming problem initially had some constructive implication, 
because in \cite{NC97} was suggested a special new kind of {\em stochastic} 
quantum networks to resolve the problem of universality. In the stochastic
{\em programmable quantum gate arrays} \cite{NC97} the size of a program is also 
only in two times bigger than the data, but the result of calculation is 
{\em non-deterministic}. The stochastic, probabilistic design described 
in many works \cite{NC97,C0,C1,B1,B4} and will not be discussed here with details. 

The models with measurements and probabilities make actual using  
programmable networks with mixed states and density matrices \cite{B2,B5}, 
but it also deserves a separate account and not presented here.  

In the present {\paper} is considered the deterministic design with pure states
and it is shown, that the problem with universality has rather formal meaning. 
From the one hand, only {\em finite} number of different operators are necessary
for {\em the universality in the approximate sense} often used by default already in 
the earliest works about the quantum networks \cite{QCN,UQC,Ek95} and so it is
possible to construct an universal (in the approximate sense) programmable 
quantum network with finite program register \cite{V1,V2,QPC02,V3}.

From the other hand, even if to use the notion of {\em the exact universality} 
\cite{Cle99} with the infinite number of programs, it is possible to
use {\em the quantum computations with continuous variables} 
\cite{LlBr98} for generalization of considered models 
for the infinite-dimensional program register \cite{QPC02}. 

The models of programmable quantum networks with pure states
considered in this {\paper} have the direct analogue with 
{\em the conditional quantum dynamics} \cite{Joz95}.

\subsubsection*{Contents}

In {\Sec{formal}} is represented {\em the formal theory of programmable 
quantum networks with pure states.} 
Definitions are revisited in {\Sec{def}}. Limitations due to laws 
of quantum mechanics and the conception of the deterministic and stochastic
programmable quantum gate arrays are discussed in {\Sec{nogo}}. Quantum
networks presented in this {\paper} are {\em deterministic} in
such classification, but some specific points relevant to the stochastic 
design may be found in {\Sec{non}}. 
The deterministic design is originated from the idea of conditional quantum
dynamics recalled in {\Sec{cond}}. In present {\paper} are also mainly are used
quantum systems with finite-dimensional Hilbert spaces, but a hybrid model 
with continuous quantum variables is also not omitted and may be found 
in {\Sec{inf}}. The design of a programmable {\sf Control-Shift} network is
represented finally in {\Sec{3bus}} and is used as a basic model for the rest 
of the {\paper}. 

In the {\Sec{models}} are discussed {\em more concrete models.}
The theory of universality in the quantum computations and control is recollected 
in {\Sec{univ}}. All necessary Hamiltonians for the programmable 
{\sf Control-Shift} network with the universal set of quantum gates are 
constructed in {\Sec{Ham}}. The relation with the theory of quantum cellular
automata is demonstrated in {\Sec{QCA}}.

\section{Formal theory of programmable quantum networks}
\label{Sec:formal}

\subsection{Definition}
\label{Sec:def}

Let us return to the main equation describing a programmable quantum network
\Eq{qprog}. It is also may be written in more general form, when
the state of the program register may be changed after the operation: 
\begin{equation}
\mathsf U \colon \bigl(\ket{D}\otimes\ket{\varPi}\bigr)
  \mapsto \bigl(\op u_\varPi\ket{D}\bigr)\otimes\ket{\varPi'}.
\tag{\ref{qprog}$'$}
\label{qprog'}
\end{equation}

\begin{figure}[htb]
\begin{center}
\IorII{\includegraphics[scale=1.25]{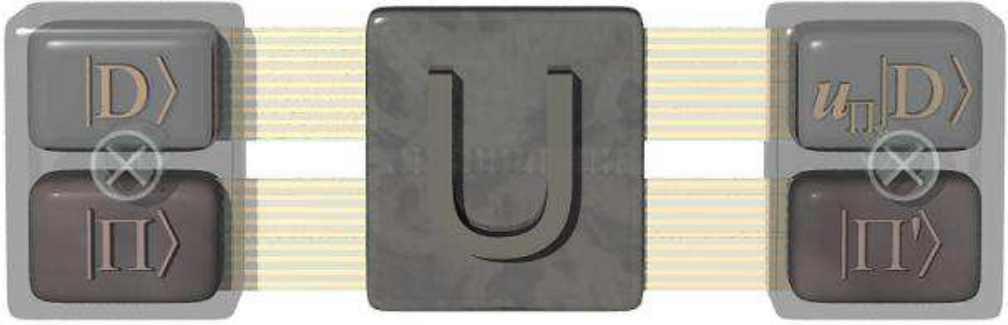}}
{\includegraphics[bb=0 0 800 300,scale=0.4]{qnet3a.jpg}}
\end{center}
\caption{Scheme of a programmable quantum network}
\label{Fig:qnet3a}
\end{figure}

The scheme of a network for such transformation is depicted on \Fig{qnet3a}.
In the \Eq{qprog'} is used expanded notation with the tensor product 
$\ket{D}\otimes\ket{\varPi}$ often omitted for simplicity
in expressions like $\ket{D}\ket{\varPi}$ or $\ket{D,\varPi}$. 

Sometime it is convenient to exchange program and data registers and
to write
\begin{equation}
\mathsf U \colon \bigl(\ket{\varPi}\otimes\ket{D}\bigr)
  \mapsto \ket{\varPi'}\otimes \bigl(\op u_\varPi\ket{D}\bigr).
\tag{\ref{qprog}$''$}
\label{qprog''}
\end{equation}

\subsection{Limitations due to quantum laws}
\label{Sec:nogo}

The \Eq{qprog} and \Eq{qprog'} was analyzed in \cite{NC97}, there
such kind of networks was called {\em the programmable quantum gate
arrays}. Sometimes they are also called {\em quantum processors}
\cite{QPC03,B1,B2,B4,B5,V2,QPC02,V3}.
In \Eq{qprog'} the state of the program register $\ket{\varPi'}$ 
after the operation should not depend on the state of the data register, 
because for two different data states it could be written
\begin{equation}
\begin{split}
\mathsf U\bigl(\ket{D_1}\otimes\ket{\varPi}\bigr) &=
  \bigl(\op u_\varPi\ket{D_1}\bigr)\otimes\ket{\varPi'_1} \\
\mathsf U\bigl(\ket{D_2}\otimes\ket{\varPi}\bigr) &=
  \bigl(\op u_\varPi\ket{D_2}\bigr)\otimes\ket{\varPi'_2}.
\label{difdat}
\end{split}
\end{equation}
The unitary operator {\sf U} does not change the scalar product and
it is possible to write
\begin{equation}
 \braket{D_1}{D_2} \underbrace{\braket{\varPi}{\varPi}}_1 = 
 \braket{D_1}{D_2} \braket{\varPi'_1}{\varPi'_2}.
\label{d1d2}
\end{equation}
If states $\ket{D_1}$ and $\ket{D_2}$ are not orthogonal 
$\braket{D_1}{D_2} \ne 0$, \Eq{d1d2} is satisfied only for
$\braket{\varPi'_1}{\varPi'_2} = 1$ and so $\ket{\varPi'_1} = \ket{\varPi'_2}$.

For $\braket{D_1}{D_2} = 0$ such an argument does not work, because \Eq{d1d2}
is always true. It is really possible to resolve \Eq{difdat} for orthogonal
data states, but in such a case for different states $\ket{\varPi'_1}$ and 
$\ket{\varPi'_2}$ due to linearity it may be written:
$$
\mathsf U \bigl((\alpha\ket{D_1}+\beta\ket{D_2})\otimes\ket{\varPi}\bigr)
  = \alpha\bigl(\op u_\varPi\ket{D_1}\bigr)\otimes\ket{\varPi'_1}
  + \beta\bigl(\op u_\varPi\ket{D_2}\bigr)\otimes\ket{\varPi'_2},
$$
so the state of program and data registers may be entangled and the equation
does not have required form.
Only for $\ket{\varPi'_1} = \ket{\varPi'_2} = \ket{\varPi'}$ the
equation may be reduced to the proper form \Eq{qprog'} for the arbitrary 
superposition of data states
$$
\mathsf U \bigl((\alpha\ket{D_1}+\beta\ket{D_2})\otimes\ket{\varPi}\bigr)
  = \bigl(\op u_\varPi(\alpha\ket{D_1}+\beta\ket{D_2})\bigr)\otimes\ket{\varPi'}.
$$

\medskip

In the \cite{NC97} also was found another important consequence of the given 
structure of \Eq{qprog'} --- {\em the orthogonality of different program states}.
Really, let as consider two different programs $\ket{\varPi}$ and
$\ket{\varXi}$. It is possible to write 
\begin{equation}
\begin{split}
\mathsf U\bigl(\ket{D}\otimes\ket{\varPi}\bigr) &=
  \bigl(\op u_{\varPi}\ket{D}\bigr)\otimes\ket{\varPi'}\\
\mathsf U\bigl(\ket{D}\otimes\ket{\varXi}\bigr) &=
  \bigl(\op u_{\varXi}\ket{D}\bigr)\otimes\ket{\varXi'}.
\label{pixi}
\end{split}
\end{equation}
Due to unitarity of {\sf U}
\begin{equation}
 \braket{\varPi}{\varXi} = 
 \bra{D}\op u_{\varPi}\op u_{\varXi} \ket{D} \braket{\varPi'}{\varXi'}
\label{scalprog}
\end{equation}
In \cite{NC97} was noted, that because only the term 
$\bra{D}\op u_{\varPi}\op u_{\varXi} \ket{D}$ in \Eq{scalprog}
depends on the state $\ket{D}$, it may be resolved only for 
$\braket{\varPi}{\varXi} = 0 = \braket{\varPi'}{\varXi'}$
{\em i.e.}, if all {\em different programs correspond to orthogonal 
states}. So the orthogonality condition \Eq{ortprog} mentioned in 
\Sec{Intro} is proved. 

\begin{Note} 
It should be mentioned, that the question about behavior of
terms like $\bra{D}\op u \ket{D}$ maybe denotes additional discussions,
but it is above a scope of presented {\paper}.
Say, it is possible to consider a real-valued analogue of the qubit  
\cite{Eucl,rebit}, a ``{\em rebit}.'' It is a vector $\ket{R}$ in 
the two-dimensional real vector space (plane). The operators $\op u$ 
now correspond to rotations of the plane. 
In such a case $\bra{R}\op u \ket{R}= \cos(\varphi)$ {\em does not depend}
on the state of the rebit (here $\varphi$ is the angle of rotation).
\end{Note}

Due to the orthogonality property \Eq{ortprog} for a program register in \Eq{qprog},
the dimension of the Hilbert space is equivalent to the number of different operators 
$\op u_\varPi$. For the exact universality we formally must have possibility to
apply the infinite number of different unitary operators $\op u$.
It was considered in \cite{NC97} as some disadvantage and it was suggested 
idea of ``probabilistic'' (non-deterministic) programmable quantum gate arrays.

\subsection{Non-unitary linear operators and non-deterministic networks}
\label{Sec:non}

It should be mentioned, that the no-go result about the exactly universal 
deterministic programmable quantum network may be considered as a more subtle 
limitation in comparison with a famous quantum no-cloning theorem \cite{noclon}.
The no-cloning theorem may be derived from the linearity of the quantum mechanics,
but the consideration above also uses the {\em unitarity}. It is essential,
because exist linear, but non-unitary operators satisfying \Eq{qprog'}
with the finite-dimensional program register.

If a data register is described by some state vector $\ket{D} \in \Hil$ with 
components $D_i$, then the minimal program register must contain the matrix of 
coefficients for the operator $A_{ij}$ and so in the simplest case such a register 
in two times bigger than the data one $\ket{A} \in \Hil \otimes \Hil$.
Here due to the usual law of composition of quantum systems for $N$-dimensional
data register such a program register is $N^2$-dimensional.

Let $\ket{A} = \sum A_{ij}\ket{i}\ket{j}$ is the state of the program
encoding the operator $\op A = \sum A_{ij}\ket{i}\bra{j}$, {\em viz} 
$
 \ket{A} = \sum_j \ket{j}\op A\ket{j} 
$
and {\sf M} is the linear non-unitary operator defined on the basis as
\begin{equation}
 \mathsf M \colon \ket{k}\ket{i}\ket{j} \mapsto \delta_{jk}\ket{i}\ket{0}\ket{0}.
\label{mulbas}
\end{equation}
Then a specific kind of \Eq{qprog'} is valid
\begin{equation}
 \mathsf M \bigl(\ket{\psi}\ket{A}\bigr) = \ket{\op A \psi}\ket{0,0},
\label{linprog}
\end{equation}
because
$
(\op A \psi)_i = \sum_j A_{ij} \psi_j = \sum_{jk}A_{ij}\delta_{jk} \psi_k.
$ 
In fact, the {\sf M} is the operator of the matrix multiplication rewritten
in a specific way.

The idea has some relation with the non-deterministic design \cite{NC97}. 
Definition of the state $\ket{U}$ representing an unitary operator $\op U$ 
may be simply changed, to ensure the unit norm
\begin{equation}
 \ket{U} = \frac{1}{\sqrt{N}}\sum_j \ket{j}(\op U\ket{j}). 
\end{equation}
For the qubit (two-dimensional Hilbert space) it coincides with the definition 
in \cite{NC97}. It is possible to rewrite $\ket{\psi}\ket{U}$ using the Bell basis 
$\Phi_\pm = \nrm2(\ket{00} \pm \ket{11})$,
$\Psi_\pm = \nrm2(\ket{01} \pm \ket{10})$
for the first two qubits \cite{NC97} 
\begin{equation}
 \ket{\psi}\ket{U} = 
 \frac{1}{2}\bigl( \ket{\Phi^+}\op U \ket{\psi} + 
  \ket{\Psi^+}\op U \op \sigma_x \ket{\psi} + 
  i\ket{\Psi^-}\op U \op \sigma_y \ket{\psi}  + 
  \ket{\Phi^-}\op U \op \sigma_z \ket{\psi}  
 \bigr). 
\end{equation}

With minimal modification such expression may be converted to
the expansion of some {\em unitary} operator 
$\mathsf{G = L + R}$, there {\sf L} is the non-unitary linear 
``programming'' operator and {\sf R} is the residue $\mathsf{R = G - L}$ 
\begin{equation}
 \mathsf G \bigl(\ket{\psi}\ket{U}\bigr) = (\op U \ket{\psi})\ket{0,0} + 
  \mathsf R \bigl(\ket{\psi}\ket{U}\bigr).
\end{equation}
So the non-unitary linear operator at least formally presents in
the model of the non-deterministic programmable quantum gate array,
but further discussion on the question is outside of the scope of this {\paper}. 

\subsection{Conditional quantum dynamics}
\label{Sec:cond}

Let us return to the programmable quantum networks denoted in \cite{NC97}
as {\em deterministic}. The result about the orthogonality of different
program states also has some positive implications and simplifies 
description. It is possible to choose the states corresponding to 
different programs as a new basis and to write instead of $\ket{\varPi_k}$ 
simply $\ket{k}$. Structure of the operator {\sf U} \Eq{qprog} becomes
quite clear in such notation, it is {\em the conditional quantum dynamics}
\cite{Ek95} described even earlier, than the programmable quantum gate
arrays mentioned above.

Such an unitary operator {\sf U} for orthogonal $\ket{k}$ from
$m$-dimensional Hilbert space may be written \cite{Ek95}
\begin{equation}
 \mathsf U = \sum_{k=0}^{m-1}  \ket{k}\bra{k}\otimes \op u_k ,
\quad
\mathsf U\bigl(\ket{k}\ket{D}\bigr) =
  \ket{k}\bigl(\op u_k\ket{D}\bigr).
\label{condU}
\end{equation}
In \Eq{condU} program and data registers are exchanged in comparison
with definition \Eq{qprog} or \Eq{qprog'} and it corresponds to 
the alternative notation \Eq{qprog''}. Such an order is used further in the
paper for convenience (see {\em Note} on the page \pageref{note:order}).

\rightpic{128bp}{
\IorII{\includegraphics[scale=0.4]{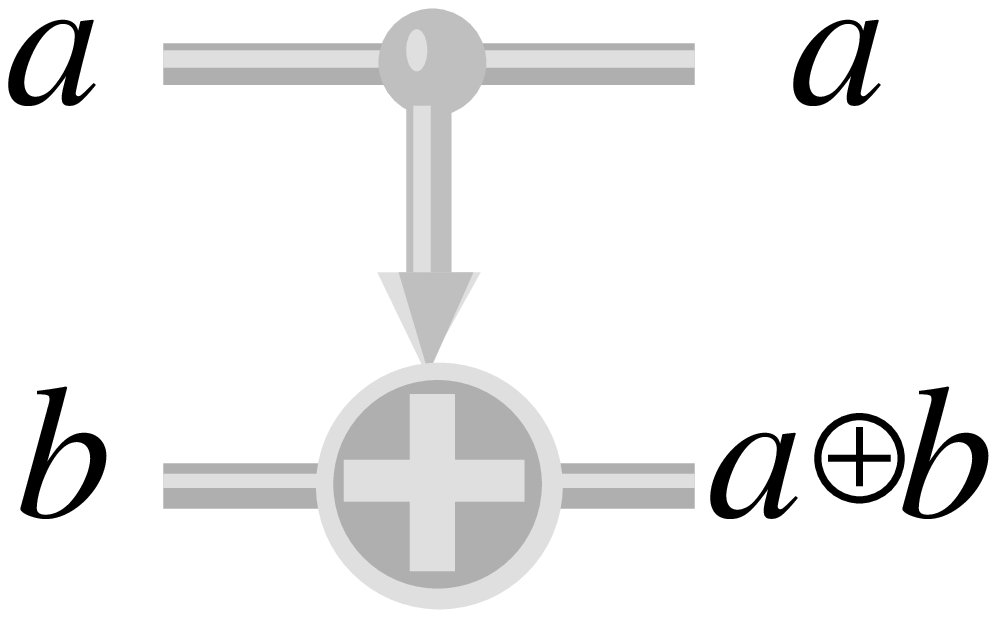}}
{\includegraphics[bb=0 0 320 200,scale=0.4]{qcnot12t.jpg}}
}{
A simplest example --- is the {\sf controlled-NOT}, {\sf c-NOT} gate. It was
used already in earliest papers about quantum computers \cite{FeyComp}
and also called {\em the measurement gate} \cite{QCN}.
It is the quantum version of a classical reversible gate with two input and 
two output states. The {\sf c-NOT} applies ${\sf NOT}$ to
a second state if and only if the first state is {\sf 1} ({\sf TRUE}), {\em i.e.},
for Boolean variables $a,b$ may be described as
$(a,b) \mapsto (a, b \xplus a)$, where ``$\xplus$'' is addition modulo 2,
$0 \xplus 0 = 1 \xplus 1 = 0$, $0 \xplus 1 = 1 \xplus 0 = 1$.
}

\begin{figure}[htb]
\begin{center}
\twox{
\IorII{\includegraphics[scale=0.4]{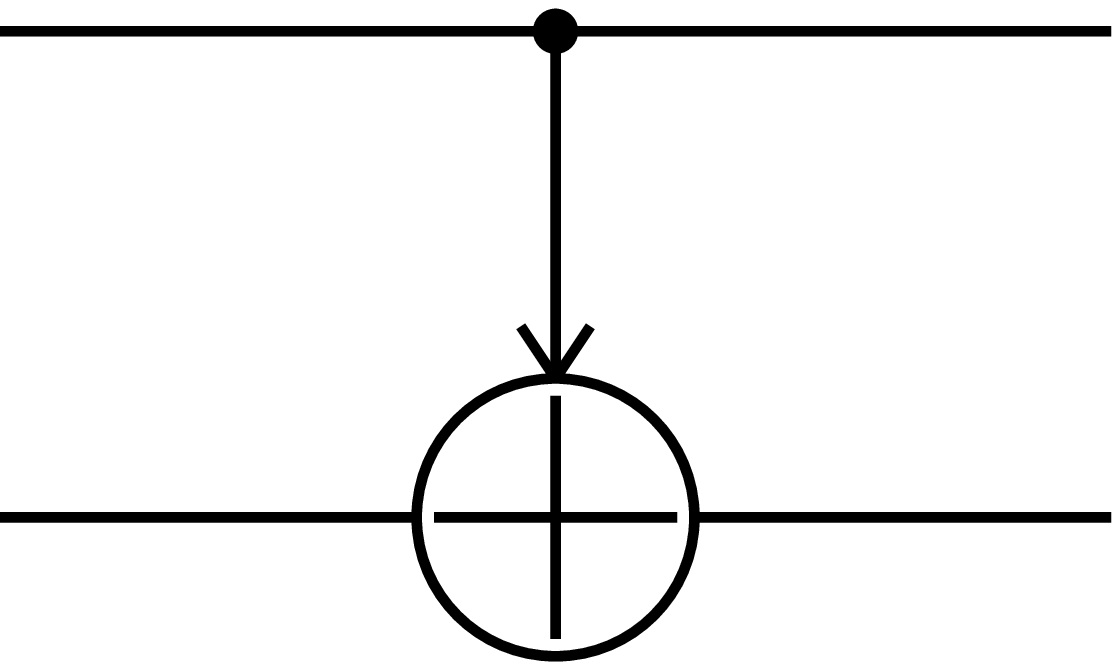}}
{\includegraphics[bb=0 0 320 240,scale=0.4]{qcnot12.jpg}}}
{a) \cNOT{12}}~
\twox{
\IorII{\includegraphics[scale=0.4,origin=c,angle=180]{qcnot.eps}}
{\includegraphics[bb=0 0 320 240,scale=0.4]{qcnot21.jpg}}}
{b) \cNOT{21}}
\end{center}
\caption{Schemes of {\sf controlled-NOT} quantum gates.  
 a) The first qubit --- is control  $\ket{a}\ket{b} \mapsto \ket{a}\ket{b \xplus a}$.
\quad b) The second qubit --- is control $\ket{a}\ket{b} \mapsto \ket{b \xplus a}\ket{b}$.
}
\label{Fig:qcnot}
\end{figure}

The quantum mechanical version is straightforward
\begin{equation}
 \cNOT{12} \colon \ket{a}\ket{b} \mapsto \ket{a}\ket{b \xplus a}.
\label{Cnot}
\end{equation}
Two possible instances of the {\sf c-NOT} gate are depicted on \Fig{qcnot}.
They also may be considered as representations of two different orders 
of program and data registers used in \Eq{condU} and \Eq{qprog}.

It is useful to consider the simple case to show principle of
matrix representation of such operators. The action of the 
{\sf c-NOT} gate for the basis may be written as
\begin{equation}
\begin{split}
\ket{0}\ket{0} &\mapsto \ket{0}\ket{0} \\
\ket{0}\ket{1} &\mapsto \ket{0}\ket{1} \\
\ket{1}\ket{0} &\mapsto \ket{1}\ket{1} \\
\ket{1}\ket{1} &\mapsto \ket{1}\ket{0}
\end{split}
\label{cnoto}
\end{equation}
so the {\sf c-NOT} gate simply exchanges two last vectors of the
basis $\ket{10} \leftrightarrow \ket{11}$ and the matrix of the
operation may be written as
\begin{equation}
\cNOT{12} =
\Mat{1&0&0&0\\
     0&1&0&0\\
     0&0&0&1\\
     0&0&1&0} 
\qquad \Col{\ket{00}\\ \ket{01}\\ \ket{10}\\ \ket{11}}
\label{cNotMat}
\end{equation}

Unlike the classical case, the difference between \cNOT{12} and \cNOT{21}
quantum gates \Fig{qcnot} is rather formal --- it is enough to change
bases for the both systems to convert one gate to another and so both qubits 
formally are equal in such operation \cite{Zur98,caus}. The new basis is 
\begin{equation}
 \ket{+} = \nrm2\,(\ket{0}+\ket{1}), \quad  \ket{-} = \nrm2\,(\ket{0}-\ket{1}).
\label{ketpm}
\end{equation}
In the new $\ket{\pm}$ basis \cNOT{12} \Eq{cnoto} is rewritten

\begin{equation}
\begin{split}
\ket{+}\ket{+}  = 
\tfrac{1}{2}\,\bigl(\ket{00}{+}\ket{01}{+}\ket{10}{+}\ket{11}\bigr) 
\mapsto&\tfrac{1}{2}\,\bigl(\ket{00}{+}\ket{01}{+}\ket{10}{+}\ket{11}\bigr)
= \ket{+}\ket{+}\\
\ket{+}\ket{-}  = 
\tfrac{1}{2}\,\bigl(\ket{00}{-}\ket{01}{+}\ket{10}{-}\ket{11}\bigr) 
\mapsto&\tfrac{1}{2}\,\bigl(\ket{00}{-}\ket{01}{-}\ket{10}{+}\ket{11}\bigr)
= \ket{-}\ket{-}\\
\ket{-}\ket{+}  = 
\tfrac{1}{2}\,\bigl(\ket{00}{+}\ket{01}{-}\ket{10}{-}\ket{11}\bigr) 
\mapsto&\tfrac{1}{2}\,\bigl(\ket{00}{+}\ket{01}{-}\ket{10}{-}\ket{11}\bigr)
= \ket{-}\ket{+}\\
\ket{-}\ket{-}  = 
\tfrac{1}{2}\,\bigl(\ket{00}{-}\ket{01}{-}\ket{10}{+}\ket{11}\bigr) 
\mapsto&\tfrac{1}{2}\,\bigl(\ket{00}{-}\ket{01}{+}\ket{10}{-}\ket{11}\bigr)
= \ket{+}\ket{-}
\end{split}
\label{cnotpm}
\end{equation}
The \Eq{cnotpm} show, that a first element is swapped if and only if
the second one is $\ket{-}$, it is just \cNOT{21} written in the new bases 
$\ket{\pm}\ket{\pm}$.
So there is no clear distinction between the control and the controlled
system. It is shown further, that such situation may be not 
valid for more difficult cases.

\smallskip

Similarly with {\sf controlled-NOT} it is possible for an arbitrary
gate $\op U= \mat{U_{00}&U_{01}\\U_{10}&U_{11}}$ 
to define the {\sf controlled-}$\op U$ gate sometimes denoted as
$\wedge_1(\op U)$ \cite{gates}
\begin{equation}
\vcenter{\hbox{
\IorII{\includegraphics[scale=0.3]{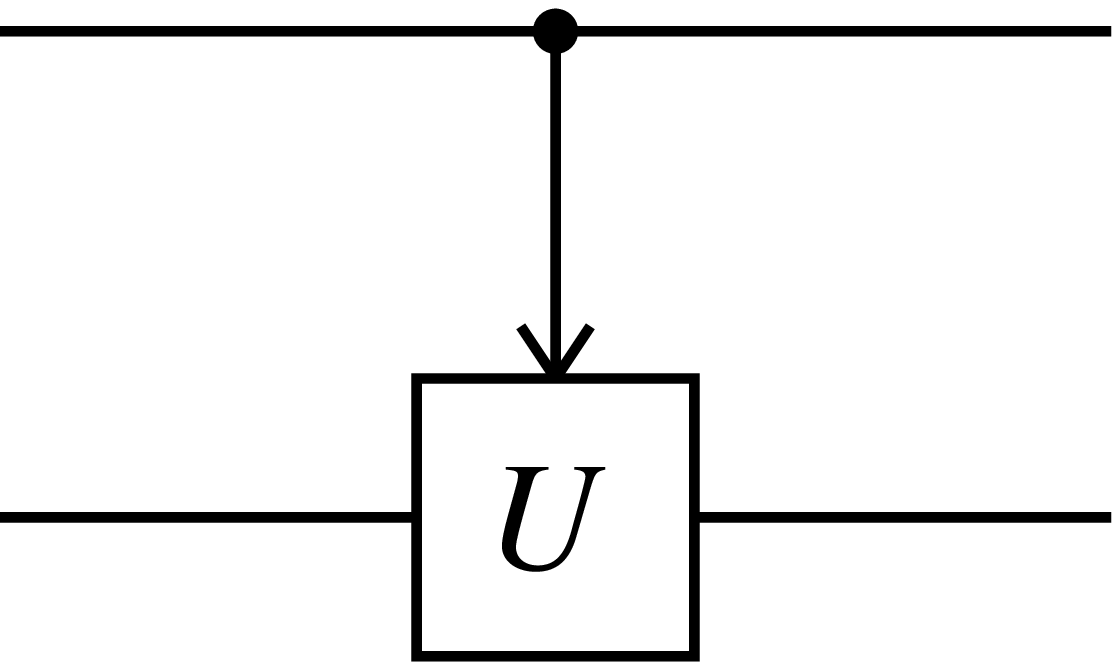}}
{\includegraphics[bb=0 0 320 240,scale=0.3]{qcU12.jpg}}%
}}\qquad
{\textstyle \bigwedge_1}(\op U) =
\Mat{1&0&0&0\\
     0&1&0&0\\
     0&0&U_{00}&U_{01}\\
     0&0&U_{10}&U_{11}} 
\qquad \Col{\ket{00}\\ \ket{01}\\ \ket{10}\\ \ket{11}}
\label{eqcU}
\end{equation}

Using the same principle, the sum used in \Eq{condU} may be written as
the block-diagonal matrix
\begin{equation}
\mathsf U = \Mat{\op u_0\\&\op u_1&&\smash{\mbox{\Huge$0$}}\\
                 &&\ddots\\\smash{\mbox{\Huge$0$}}&&&\op u_{m-1}}.
\label{UMat}
\end{equation} 

\begin{Note}\label{note:order}
For the $m$-dimensional program register and the $n$-dimensional data
register such $\mathsf U$ is \mbox{$mn \times mn$} matrix with
only nonzero elements are $n \times n$ blocks with matrices $\op u_k$ 
on diagonal of  $\mathsf U$. The understanding notation \Eq{UMat}
may be used if the program register is the first system like in \Eq{qprog''}
and \Eq{condU}.
\end{Note}

Due to the orthogonality condition \Eq{ortprog} the dimension of the
program register often much bigger $m \gg n$ or maybe even infinite and so, 
unlike the example with {\sf c-NOT} gate, two systems used for data
and program registers in general are not equivalent.

\subsection{Infinite-dimensional program register}
\label{Sec:inf}

It was already mentioned, that for the universality in the exact sense 
the dimension of a program register should be infinite \cite{NC97}. 
Let us consider such a case. Such a programmable network uses
both continuous and discrete quantum variables for the program
and the data registers respectively \cite{QPC02} and so may be considered
as an example of the {\em hybrid} quantum network \cite{Llo00}. 

The system is {\em hybrid} also in other meaning \cite{hybr2}, 
{\em i.e.}, it is the possible approach to the unified description of 
a quantum system controlled by some analogue parameters. 
It make some bridge with usual (non-programmable) quantum networks 
based on pseudo-classical description of ``tuneable'' gates 
depicted schematically on \Fig{qnet12}b.

For transition to the continuous variables it is possible to change 
the sum in \Eq{condU} to the integral and write formally \cite{QPC02}
\begin{equation}
 \mathsf U = \int \bigl(\ket{q}\bra{q} \otimes \op u(q)\bigl) dq ,
 \quad
\mathsf U\bigl(\ket{q}\ket{D}\bigr) =
  \ket{q}\bigl(\op u(q)\ket{D}\bigr).
\label{contU}
\end{equation}
The \Eq{contU} describes the family of gates (matrices) $\op u(q)$
parameterized by some continuous variable $q$. Elements of the basis 
in infinite-dimensional Hilbert space are denoted as $\ket{q}$.

A standard example --- is the space of functions  $\psi(x)$ on a line 
and Dirac delta functions as the basis 
\begin{equation}
\ket{q} = \delta(x-q), \quad \braket{q}{\psi(x)} = \psi(q).
\label{cord}
\end{equation}

The sign $\otimes$ of tensor product may be skipped in the expression 
like \Eq{contU}, because for such a hybrid case with continuous and 
discrete quantum variables there is simple interpretation. The
tensor product of the space of functions on a line and some 
finite-dimensional vector space may be represented as a space
of multi-component functions on the line with values in the
vector space.
Say the both program and data registers may be encoded in
a wave vector $\ket{\Psi(x)}$ of one particle. 

\begin{figure}[htb]
\begin{center}
\IorII{\includegraphics[scale=1.25]{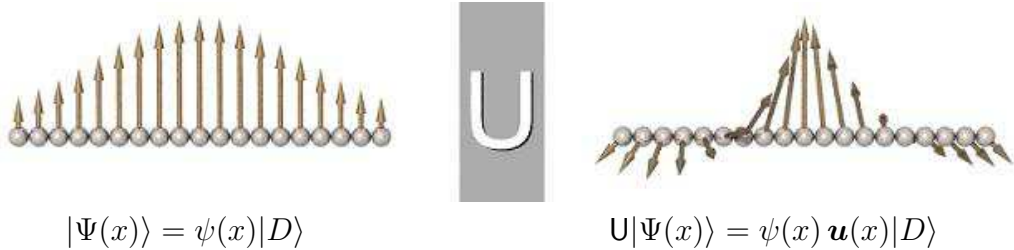}}
{\includegraphics[bb=0 0 800 150,scale=0.4]{distrib.jpg}}

$\ket{\Psi(x)} = \psi(x)\ket{D}$ \hfil\hfil
$\mathsf U\ket{\Psi(x)}=\psi(x)\, \op u(x) \ket{D} $

\end{center}
\caption{A scheme of the process \Eq{Upsi} with a ``distributed'' qubit.}
\label{Fig:distr}
\end{figure}

The space of linear operators in such a case
may be represented as a space of matrix-valued operators on
the line. For example \Eq{contU} may be expanded as
\begin{equation}
\mathsf U\bigl(\psi(x)\ket{D}\bigr) =
 \int \delta(x-q) \psi(q) \op u(q) \ket{D} dq =
 \psi(x) \op u(x) \ket{D}.
\label{Upsi}
\end{equation}
Using the example with one particle, \Eq{Upsi} may be considering
as a process with ``twisting along $x$'' of the wave function 
$\ket{\Psi(x)}$ depicted on \Fig{distr}.

It is also possible to use the momentum basis of the periodic functions 
\begin{equation}
\widetilde{\ket{p}} = \exp(2\pi i\,p\,x).
\label{imp}
\end{equation}
With the new basis it is possible to write
\begin{equation}
 \tilde{\mathsf{U}} = \int \bigl(\widetilde{\ket{p}}\widetilde{\bra{p'}} \op u(p)\bigl) dp ,
 \quad
\tilde{\mathsf{U}}\bigl(\widetilde{\ket{p}}\ket{D}\bigr) =
  \widetilde{\ket{p'}}\bigl(\op u(p)\ket{D}\bigr),
\label{momU}
\end{equation}
there is also taken into account the possibility of change of the momentum $p \to p'(p)$ 
(the program register) similarly with \Eq{qprog'}.

\begin{figure}[htb]
\begin{center}
\IorII{\includegraphics[scale=1.25]{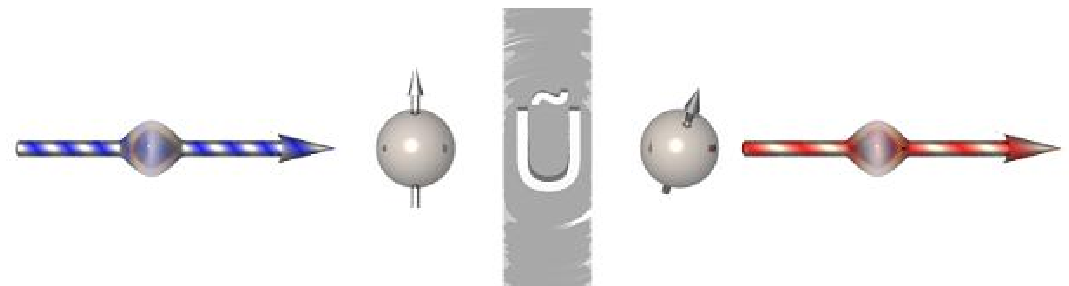}}
{\includegraphics[bb=0 0 800 200,scale=0.4]{scatt.jpg}}

$t_1$ \hfil  $t_2$

\end{center}
\caption{Illustration of \Eq{momU} with scattering process.
}
\label{Fig:scatt}
\end{figure}

More details may be found in \cite{QPC02}, let us only consider a visual example
with ``a scattering process'' \Fig{scatt}, there are used two different systems.
 
Before the scattering (\Fig{scatt}, time $t_1$) the first system was described by 
a ``monochrome'' wave function $\exp(i k x)$ and the second one --- is a
localized qubit $\ket{D}$. After the ``inelastic'' scattering process 
(\Fig{scatt}, time $t_2$) the first system may have a state with other $k' = k'(k)$ 
and the qubit changed a state to $\op u(k)\ket{D}$.

Such a scattering process maybe too formal and abstract, but it provides
the understanding analogue between the programmable quantum network and
a more traditional design with a quantum system controlled by some 
external devices like lasers.

It should be mentioned, that the considered model with continuous variables 
also provides some link with the stochastic programmable quantum gate arrays. 
It was already mentioned, that such a model produces a correct answer with some
probability. For $N$-dimensional data register the probability is
$1/N$, {\em i.e.},  $2^{-n}$ for a system with $n$ qubits \cite{NC97}.
For some particular case of encoding $U(1)$ operations, 
it is possible to make the probability higher for the bigger program 
register with approach to the unit probability of success in the infinite
limit \cite{C0,C1}.  

On the other hand, for the deterministic design it is also possible to consider
the infinite-dimensional program register as a limit of finite-dimensional one.
Sequences of finite-dimensional networks used for such limits have quite 
different properties for stochastic and deterministic approach. For the first 
one we may apply an arbitrary operation $U(1)$, similar with rotations on 
arbitrary angles $2 \pi \phi$, but it succeeds only with some probability
increasing with approach to the infinite limit.

For the second one, the operation is always successful, but we may apply 
only finite number of different operations $U(1)$, {\em i.e.}, rotations on some
fixed angles $2 \pi k/N$, and for the infinite limit the rotations
cover the full circle. 
Despite of such difference for the finite case, {\em the infinite limits for both
deterministic and stochastic networks for $U(1)$ operation are essentially 
the same}. A rather technical proof of the interesting fact may be found in 
\cite{QPC02} and is not reproduced here.

\subsection{Programmable quantum networks with three buses}
\label{Sec:3bus}

It should be mentioned, that \Eq{qprog} and all derivative equations
considered here are still do not have the complete analogue with the idea
of usual (von Neumann) computer architecture, because formally they are 
describe {\em only one step} of a program.

From the one hand, such a picture sometimes may be appropriate for
the theory of quantum computing. On the other one, it is very limiting,
because even in the standard definition of the universal set of quantum
gates, it is used {\em the arbitrary composition} of 
the gates from the set \cite{QCN,UQC,Ek95,Cle99,gates}.

\begin{figure}[htb]
\begin{center}
\IorII{\includegraphics[scale=1.25]{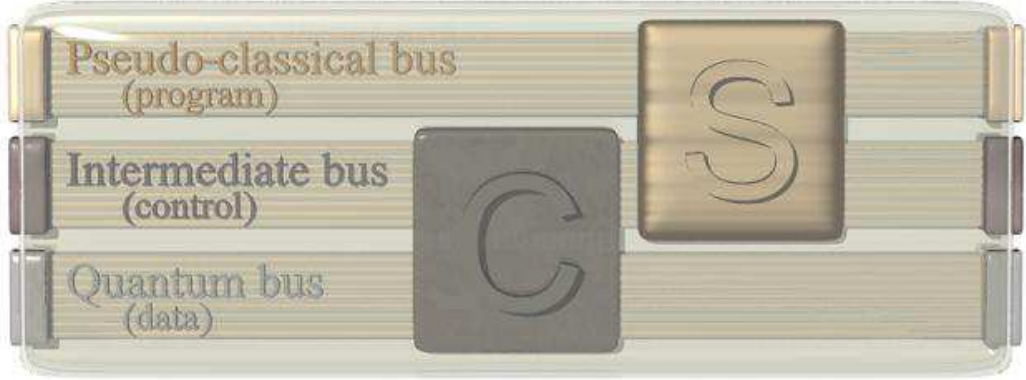}}
{\includegraphics[bb=0 0 800 300,scale=0.4]{qproc.jpg}}
\end{center}
\caption{The programmable quantum network (processor) with three buses}
\label{Fig:qproc}
\end{figure}

Such an idea of the universality may be implemented also for the programmable
quantum networks with pure states \cite{V1,V2,QPC02,V3} discussed
in this {\paper}. The scheme \Fig{qproc} uses two operators
and three buses.

The {\sf C}, {\sf Control} --- is the already considered
operator \Eq{qprog''} of {\em one computational step} with 
second (control) and third (quantum data) buses. The {\sf S}, {\sf Shift} 
operator must change the state $\ket{k}$ of a control bus after each step
using first (program) and second buses. 

The buses also may be called {\em pseudo-classical}, 
{\em intermediate} and {\em quantum}, because the quantum bus may 
contain arbitrary superposition of states, the intermediate bus
should use only orthogonal states, but it is still may
not be considered by an entirely classical way, because it is linked
via {\sf C} operator with the quantum bus and the specific character
of such a design was already discussed above. But already for 
description of the pseudo-classical bus and the operator {\sf S}
it is at least formally enough to use the theory of reversible 
classical gates.

In principle, {\sf S} may correspond to an arbitrary {\em reversible}
classical program, but here is convenient to consider the simplest
case with the {\em cyclic shift}. Let us suggest, that a data register
is described by $m$-qubits, there is a finite set with 
$N=2^n$ quantum gates, and for our purposes it is necessary to apply
a sequence with $L$ gates from the given set. For example we may consider
the finite set of universal gates and the task of the approximation of 
an arbitrary gate with the necessary precision using up to $L$ gates.

For the design with the cyclic shift operator {\sf S} a program register
must have size $(L-1)n$ qubits and acts on $L\,n$ qubits of
program and control registers as
\begin{equation}
 \mathsf S \colon  \ket{k_L,k_{L-1},\ldots,k_2}\ket{k_1} 
           \mapsto \ket{k_1,k_L,\ldots,k_3}\ket{k_2}.
\label{Shft}
\end{equation}
Action of {\sf C} was already defined in \Eq{condU} 
\begin{equation}
\mathsf C \colon \ket{k}\ket{D} \mapsto
  \ket{k}\bigl(\op u_k\ket{D}\bigr).
\tag{\ref{condU}$'$}
\label{Ctrl}
\end{equation}
and so the composition of {\sf C} and {\sf S} may be written
\begin{equation}
 \mathsf{S\, C}\Bigl(\ket{k_L,k_{L-1},\ldots,k_2}\ket{k_1}\ket{D}\Bigr) 
           = \ket{k_1,k_L,\ldots,k_3}\ket{k_2}\bigl(\op u_k\ket{D}\bigr).
\label{compCS}
\end{equation}
It is only one step. Let us denote the state of program and control registers 
as $\ket{K}=\ket{k_L,k_{L-1},\ldots,k_2}\ket{k_1}$. 
For $L$ steps $\ket{K}$ returns to the initial state and so it is
possible to write \cite{V1,V2,QPC02,V3}
\begin{equation}
 (\mathsf{S\, C})^L \colon \ket{K}\ket{D}
 \mapsto \ket{K} \bigl(\op u_L \cdots \op u_2\op u_1 \ket{D}\bigr)
\label{cyCS}
\end{equation}

\begin{figure}[htb]
\begin{center}
\IorII{\includegraphics[scale=0.6]{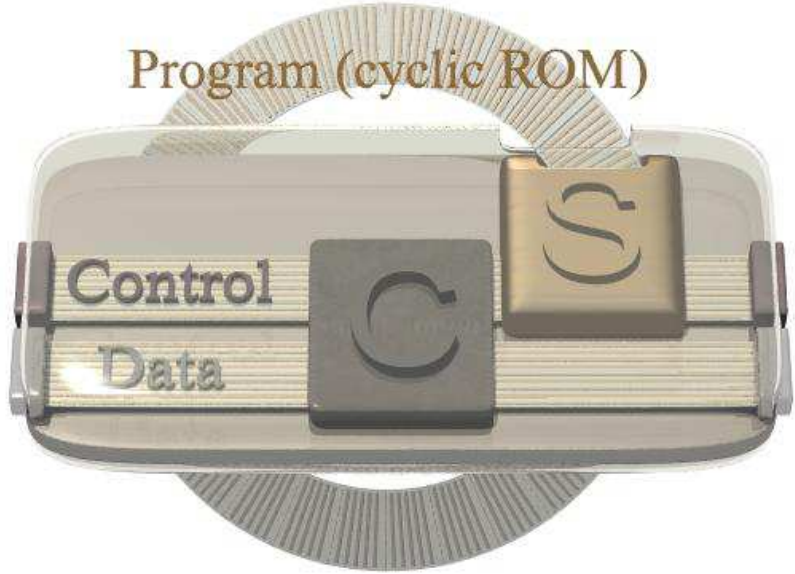}}
{\includegraphics[bb=0 0 512 384,scale=0.4]{qprocyc.jpg}}
\end{center}
\caption{A scheme of {\sf Control-Shift} network}
\label{Fig:qprocyc}
\end{figure}

Here the program bus should be rather compared with the {\em cyclic read-only memory}
({\sf ROM}). A scheme of the programmable ``{\sf Control-Shift}'' network with 
the cyclic shift register is depicted on \Fig{qprocyc}.

\section{Models of programmable quantum networks}
\label{Sec:models}

\subsection{Universal sets of quantum gates}
\label{Sec:univ}

\begin{figure}[htb]
\begin{center}
\IorII{\includegraphics[scale=0.6]{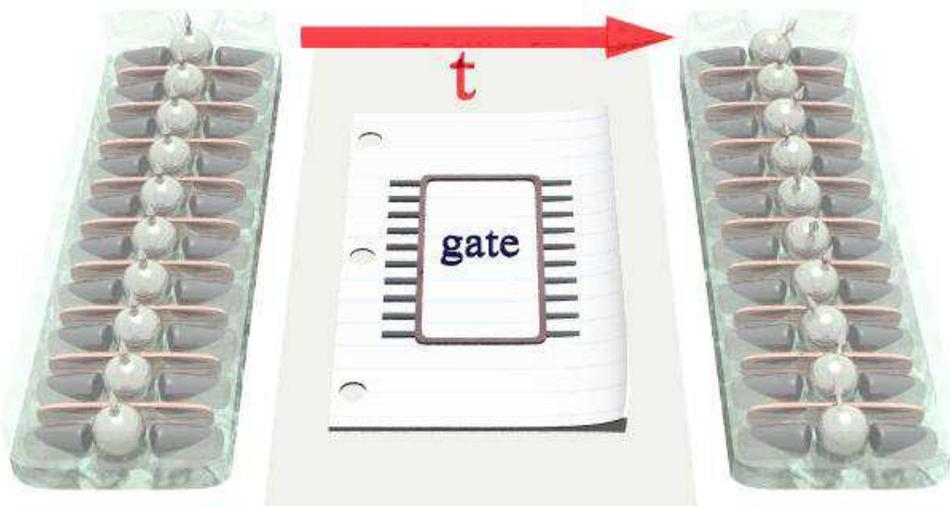}}
{\includegraphics[bb=0 0 640 320,scale=0.4]{spins01.jpg}}
\end{center}
\caption{A scheme of a gate and evolution of a quantum system}
\label{Fig:evolnet}
\end{figure}

In the theory represented above were used rather abstract models. 
Standard relation between the scheme of a quantum gate and the evolution of 
the quantum system is depicted on \Fig{evolnet}.
The input and output of the network --- is really the same system at
two different moments of time and the gate represents change of the state 
of the system during given period due to some interactions.

Idea of decomposition using the universal set of quantum
gates was already few times mentioned in the present {\paper}, {\em e.g.},
it is relevant to the model of the {\sf Control-Shift} network 
described by \Eq{cyCS}.
Let us recall the basic principles of the theory of the universality
in quantum computations and control.

There is so-called {\em infinitesimal} approach with the Lie algebras 
\cite{Ek95,DV95,UnSim}. It is convenient also due to the direct relation 
with Hamiltonians of quantum systems. For the system with the constant 
Hamiltonian $\op H$ the evolution during the time $t$ is described by 
the unitary operator (gate)
\begin{equation}
 \op U = \op U(\tau) = \exp(-i \, \op H \, \tau).
\label{eHt}
\end{equation}
If two Hamiltonians $\op H_1$, $\op H_2$ correspond to gates 
$\op U_1$, $\op U_2$, then 
\begin{equation}
 \op U_1 \op U_2 = \exp(-i \, \op H_1 \tau)  \exp(-i \, \op H_2 \tau)
 =  \exp\bigl(-i \, (\op H_1+\op H_2)\,\tau \bigr) + O(\tau^2),
\label{sumH} 
\end{equation}
there $O(\tau^2)$ is an error of order $\tau^2$ and so for small $\tau$
the sum of Hamiltonians corresponds to the composition of the gates.
It is also quite standard to use an expression for commutators 
\cite{Ek95,DV95,UnSim}
\begin{equation}
 \op U_1 \op U_2 \op U_1^{-1} \op U_2^{-1} 
 = \exp\bigl(-(\op H_1\op H_2-\op H_2\op H_1) \tau^2 \bigr) + O(\tau^3) 
\label{comH}
\end{equation}
and so such a product with four terms is approximately equal to the action of
a gate with a Hamiltonian $-i[\op H_1,\op H_2]$ and a parameter $\tau^2$, 
{\em i.e.}, if to consider $\tau = \sqrt{t}$ the \Eq{comH} has
precision $O(t^{1.5})$.

Due to \Eq{sumH} and \Eq{comH}, which valid for infinitesimal values 
$\tau \to 0$, it is possible to formulate a condition of universality
using Hamiltonians \cite{Ek95,DV95,UnSim}. The set of Hamiltonians 
$\op H_k$ corresponds to the universal set of gates, if it is possible to produce 
an arbitrary Hamiltonian using linear combinations of $\op H_k$ 
and they commutators of any order, {\em i.e.}, 
$i \bigl[\op H_j,i[\op H_k, \ldots \,]\bigr]$.

Such definition is convenient for a test of universality, but for
concrete tasks it may produce a problem due to equations for commutators.
Already for \Eq{comH} with $\tau^2$ it is not clear, how to
choose a parameter $\tau$ for the generation of a gate with a good precision, 
and the problem even worst for commutators of $k$-th order with $\tau^k$.

There is a way to produce an expression with the first degree of $\tau$
instead of $\tau^2$ in \Eq{comH} and without error at all
if to use special choice of Hamiltonians \cite{Clif}. Let us
consider a system with $n$-qubits. As a basis in the space of
Hermitian matrices (Hamiltonians) may be used $4^n$ different
tensor products of four Pauli matrices
\begin{equation}
 \op H_{\bf j} = 
 \op\sigma_{j_1} \otimes \op\sigma_{j_2} \otimes \cdots \otimes \op\sigma_{j_n},
 \quad j_k = 0,\ldots,3
\label{prtH}
\end{equation}
where $\mathbf j = (j_1,j_2,\ldots,j_n)$ is notation for multi-index and
\begin{equation}
 \op\sigma_0 = \Mat{1&0\\0&1}\!,~
 \op\sigma_1 = \Mat{0&1\\1&0}\!,~
 \op\sigma_2 = \Mat{0&i\\-i&0}\!,~
 \op\sigma_3 = \Mat{1&0\\0&-1}\!.
\label{sigmas}
\end{equation}
It is also widely used the alternative notation:  
$\op 1$,  $\op\sigma_x$, $\op\sigma_y$, $\op\sigma_z$ respectively.

All Hamiltonians \Eq{prtH} have properties
\begin{equation}
  \op H_{\bf j} \op H_{\bf k}= \pm \op H_{\bf k} \op H_{\bf j},
\quad \op H_{\bf j}^2 = \op 1. 
\label{clcom}
\end{equation}
Unity of the square in \Eq{clcom} ensures a simple expression for the exponent 
\begin{equation}
\exp(i \phi \op H_{\bf k}) = \cos(\phi)\op 1 + i\sin( \phi) \op H_{\bf k}
\label{expH}
\end{equation}
and using the identity 
$$
\exp(\op A)\exp(\op B)\exp(-\op A) = 
\exp\bigl(\exp(\op A)\op B\exp(-\op A)\bigl)
$$
it is possible to find the precise expression for the exponent of commutator
\begin{equation}
e^{-[\op H_{\bf j},\op H_{\bf k}] \tau} =
\begin{cases}
\op 1, & \op H_{\bf j}\op H_{\bf k} = +\op H_{\bf k}\op H_{\bf j} \\
e^{i \frac{\pi}{4} \op H_{\bf j}}\,e^{2 i \tau \op H_{\bf k}}\,
e^{-i \frac{\pi}{4} \op H_{\bf j}},
& \op H_{\bf j}\op H_{\bf k} = -\op H_{\bf k}\op H_{\bf j}
\end{cases}.
\label{comclH}
\end{equation}
Unlike \Eq{comH} the \Eq{comclH} contains the first degree of $\tau$
and for such a method of generation of arbitrary gates there are 
only errors related with \Eq{sumH}, which have order $O(t^2)$.

It is well-known that one- and two-gates are enough for universality 
\cite{Ek95,gates,DV95}. For the work with Hamiltonians like \Eq{prtH} with
properties \Eq{clcom} it is convenient to use an universal set with 
$2n+1$ elements 
\cite{Clif}
\begin{subequations}\label{gZXD}
\begin{eqnarray}
 \op Z_0  &=& \op\sigma_3^{(1)} = 
 \op\sigma_3\otimes\underbrace{\op\sigma_0\otimes\cdots\otimes\op\sigma_0}_{n-1}\, , 
 \label{gZ0} \\
 \op Z_1 &=& \op\sigma_3^{(2)} = 
 \op\sigma_0\otimes\op\sigma_3\otimes
 \underbrace{\op\sigma_0\otimes\cdots\otimes\op\sigma_0}_{n-2}\, , 
 \label{gZ1}\\
  \op X_k &=& \op\sigma_1^{(k+1)} =
  {\underbrace{\op\sigma_0\otimes\cdots\otimes\op\sigma_0}_k\,}\otimes
 \op\sigma_1\otimes\underbrace{\op\sigma_0\otimes\cdots\otimes\op\sigma_0}_{n-k-1}\, , 
 \label{gXk}\\
  \op D_k &=& \op\sigma_3^{(k+1)}\op\sigma_3^{(k+2)} =
  {\underbrace{\op\sigma_0\otimes\cdots\otimes\op\sigma_0}_k\,}\otimes
 \op\sigma_3\otimes\op\sigma_3\otimes
 \underbrace{\op\sigma_0\otimes\cdots\otimes\op\sigma_0}_{n-k-2}\, . 
 \label{gZZk}
\end{eqnarray}
\end{subequations}
Really the set of gates differs from suggested in \cite{Clif} on nonessential
change of the basis ($\op\sigma_1 \leftrightarrow \op\sigma_3$). After such exchange
all two-qubit gates have diagonal form due to \Eq{gZZk}. 
It should be reminded, that the gates are generated using \Eq{eHt} with given 
Hamiltonians \Eq{gZXD}. It may prevent some confusion, because all operators in 
form \Eq{prtH} are not only Hermitian, but also are unitary and so they are 
widely used as gates in many works.

\smallskip

There is also alternative approach to the universality, then instead of
consideration of Hamiltonians and infinitesimal (small) transformations,
it is considered a question, how to decompose precisely the given unitary 
matrix (with $4^n$ parameters for $n$ qubits) on product of one 
and two-qubit gates. Standard choice is: all one-qubit gates together with
{\sf c-NOT} gate \cite{tuc,nist,hels}, {\em e.g.}, in \cite{hels}
is suggested an algorithm for the decomposition with $4^n$ one-qubit gates 
and $4^n-2^n$ {\sf c-NOT} gates. 

It should be mentioned, that \cNOT{12} gate also may be transformed 
to diagonal form, if to change the basis of the {\em second} qubit 
to $\ket{\pm}$ \Eq{ketpm}. 
Such transformation is described by the Hadamard matrix 
$\op H = \nrm{2}\mat{1&1\\1&-1}$ 
and already was used above on page~\pageref{ketpm}, there it
was applied to {\em both} qubits for the transition
between \cNOT{12} and \cNOT{21} gates.
After such diagonalization the Hamiltonian \Eq{cNotMat} may be written
as
\begin{equation}
\text{\sf d-NOT} = {\textstyle\bigwedge}_1(\op\sigma_3) =
\Mat{1&0&0&0\\
     0&1&0&0\\
     0&0&1&0\\
     0&0&0&-1} 
\label{dNotMat}
\end{equation}
Really, any {\sf controlled-$\op U$} gate \Eq{eqcU} may be
transformed to diagonal form by some change of a basis, {\em i.e.},
using one-qubit operations with the second qubit. It is simply a basis
there $\op U$ itself is diagonal.

\subsection{Hamiltonians of elementary programmable gates}
\label{Sec:Ham}
\subsubsection{Control gates}

Let us consider models of basic quantum gates, necessary for  
implementation of the universal programmable networks. In \Sec{univ} 
it was shown, that it is enough to use ``powers'' of Pauli matrices 
\begin{equation}
\exp(i \op\sigma_k \tau)  = \cos(\tau)+ i\sin(\tau)\op\sigma_k
\label{expsig}
\end{equation}
together with some simple diagonal two-qubit gate. In the programmable
quantum networks are used controlled gates and so basic elements are
two- and three-gates.

The Hamiltonian of controlled versions of a gate may be found 
using the simple expression
\begin{equation}
e^{i{\textstyle\mat{\op 0 & \op 0 \\ \op 0 & \op H}} \tau} =
\Mat{\op 1 & \op 0 \\ \op 0 & e^{i \op H \tau}},
\label{expcH}
\end{equation}
where $\op 0$, $\op 1$ are zero and identity matrices respectively.
So Hamiltonians of two-gates for the controlled one-qubit operations used in 
the universal set \Eq{gZXD} may be represented as
\begin{equation}
 \op{h_X^{\sf c}} = \Mat{0&0&0&0\\0&0&0&0\\0&0&0&1\\0&0&1&0},
\quad
 \op{h_Z^{\sf c}} = \Mat{0&0&0&0\\0&0&0&0\\0&0&1&0\\0&0&0&-1}.
\label{HctrlXZ}
\end{equation}
The three-qubits Hamiltonian for the controlled diagonal gates \Eq{gZZk} is
\begin{equation}
\op{h_{D}^{\sf c}} = \Mat{
0&0&0&0&0&0&0&0\\
0&0&0&0&0&0&0&0\\
0&0&0&0&0&0&0&0\\
0&0&0&0&0&0&0&0\\
0&0&0&0&1&0&0&0\\
0&0&0&0&0&-1&0&0\\
0&0&0&0&0&0&-1&0\\
0&0&0&0&0&0&0&1}.
\label{HctrlD}
\end{equation}
Let us show, how to use instead of \Eq{HctrlD} only two-qubit Hamiltonians.
\begin{multline}
\op{h_{D}^{\sf c}} = \ket{1}\bra{1} \otimes \op\sigma_z \otimes \op\sigma_z =
\frac{1}{2}(\op 1 - \op\sigma_z) \otimes \op\sigma_z \otimes \op\sigma_z \\
 = \frac{1}{2}\bigl(\op\sigma_z^{(2)}\op\sigma_z^{(3)} 
               -\op\sigma_z^{(1)}\op\sigma_z^{(2)}\op\sigma_z^{(3)}\bigr) 
 = \frac{1}{2}\bigl(\op\sigma_z^{(2)}\op\sigma_z^{(3)} -
    \frac{i}{2}[\op\sigma_x^{(1)}\op\sigma_z^{(2)},
                \op\sigma_y^{(1)}\op\sigma_z^{(3)}]\bigr).
\label{decHd}
\end{multline}
If to consider the three-qubit gate generated by $\op{h_{D}^{\sf c}}$,
due to the decomposition \Eq{decHd} and \Eq{comclH} it is possible to write
\begin{multline}
\op{U_{D}^{\sf c}}(\tau)=
\exp(i\op{h_{D}^{\sf c}}\tau) =
  \exp(i\frac{\tau}{2}\op\sigma_z^{(2)}\op\sigma_z^{(3)})
 \exp(-i\frac{\tau}{2}\op\sigma_z^{(1)}\op\sigma_z^{(2)}\op\sigma_z^{(3)})\\
~ = \exp(i\frac{\tau}{2}\op\sigma_z^{(2)}\op\sigma_z^{(3)})
\, \exp(i\frac{\pi}{4}\op\sigma_x^{(1)}\op\sigma_z^{(2)})
\, \exp(-i\frac{\tau}{2}\op\sigma_y^{(1)}\op\sigma_z^{(3)})
\, \exp(-i\frac{\pi}{4}\op\sigma_x^{(1)}\op\sigma_z^{(2)}).
\label{decUd}
\end{multline}

\medskip

Let us consider also a Hamiltonian with two control qubits for 
implementation of all three Pauli matrices 
\begin{equation}
\op{h_{P}^{\sf c}} = \Mat{
0&0&0&0&0&0&0&0\\
0&0&0&0&0&0&0&0\\
0&0&0&1&0&0&0&0\\
0&0&1&0&0&0&0&0\\
0&0&0&0&0&i&0&0\\
0&0&0&0&-i&0&0&0\\
0&0&0&0&0&0&1&0\\
0&0&0&0&0&0&0&-1}.
\label{HctrlXYZ}
\end{equation}

It is the simple example of a Hamiltonian for the programmable network with 
four different ``programs'' and one data qubit. If first two qubits 
have the state $\ket{0}\ket{0}$, the data qubit is not changed, but if they have state
$\ket{0}\ket{1}$, $\ket{1}\ket{0}$, or $\ket{1}\ket{1}$ during a period
$\tau$, then the data qubit is changed as 
$\ket{D} \mapsto \exp(i \op\sigma_k \tau)\ket{D}$ 
with $k=1,2,3$ respectively.

From the one hand, the example shows a specific kind of the exactly universal
programmable quantum gate satisfying \Eq{condU} with the finite-dimensional control register. 
Here the continuous parameter $\tau$ is used to implement the infinite number
of different programs. From the other hand, for a programmable quantum
network it is more appropriate to fix some period $\Delta\tau$ and 
use consequent application of gates $\op u_k = \exp(i \op\sigma_k \Delta\tau)$ 
via the three-buses design discussed in \Sec{3bus}. 

Here is not suggested, that $\Delta\tau$ should be infinitesimally small.
Using the infinitesimal parameter is not the only way of approach 
to a continuous limit. Most Hamiltonians used here
generate gates with a simple periodic behavior due to \Eq{expH} and 
if to choose $\Delta\tau$ as an {\em irrational} multiple of $\pi$, 
it is possible to approximate any real parameter with an arbitrary
precision. Such an idea really was used already in the earliest
works about the theory of universal quantum computations \cite{QCN,UQC,Ek95}.

\medskip

Using qubits for the control is convenient, but is not necessary. {\em E.g.},
it is enough for universality to use only two Pauli matrices, and
so instead of two qubits it is possible to use one {\em qutrit}
for `$3{\to}2$' control, {\em i.e.}, a quantum system with three states 
$\ket{0}$,$\ket{1}$ and $\ket{2}$ 
and to write a Hamiltonian like
\begin{equation}
\op{h_{\rhd}^{\sf c}} = \Mat{
0&0&0&0&0&0\\
0&0&0&0&0&0\\
0&0&0&1&0&0\\
0&0&1&0&0&0\\
0&0&0&0&1&0\\
0&0&0&0&0&-1} \qquad
\Col{\ket{0}\ket{0}\\\ket{0}\ket{1}\\
     \ket{1}\ket{0}\\\ket{1}\ket{1}\\
     \ket{2}\ket{0}\\\ket{2}\ket{1}}.
\label{Hctrit}
\end{equation} 

\begin{figure}[htb]
\begin{center}
\IorII{\includegraphics[scale=0.6]{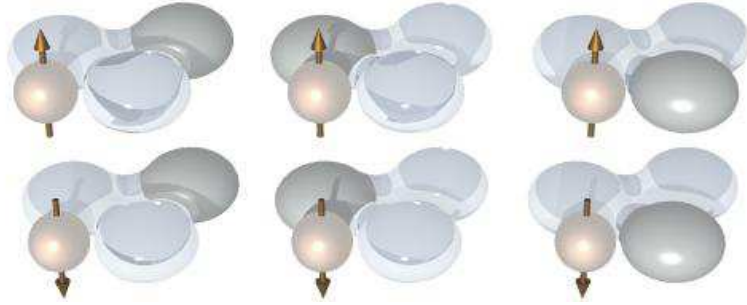}}
{\includegraphics[bb=0 0 512 192,scale=0.4]{qu3x2.jpg}}
\end{center}
\caption{A scheme of six basic states for `$3{\to}2$' control \Eq{Hctrit}. The qubit
is depicted as a system with spin and the qutrit as a ``triple quantum dot.''}
\label{Fig:qu3x2}
\end{figure}

The very rough scheme of such a system is depicted on \Fig{qu3x2}. Here a 
qubit --- is a system with spin-$1/2$, and a qutrit is represented as a 
quantum system distributed in a potential of a triple quantum dot ``molecule.'' 
It is suggested, that in the state $\ket{0}$ the qutrit is not interact 
with the qubit, but the states $\ket{1}$ and $\ket{2}$ already affect 
on the spin system and produce some evolution of the qubit
like $\ket{D} \mapsto \exp(i \op\sigma_k \tau)\ket{D}$, $k=1,2$.

\begin{figure}[htb]
\begin{center}
\IorII{\includegraphics[scale=0.6]{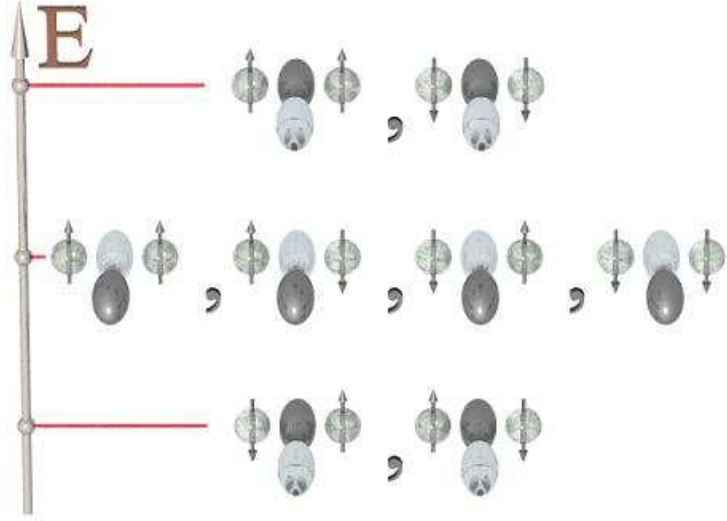}}
{\includegraphics[bb=0 0 512 384,scale=0.4]{qudcd.jpg}}  
\end{center}
\caption{A scheme of energy levels for the diagonal Hamiltonian of control 
 \Eq{EctrlD}}
\label{Fig:qudcd}
\end{figure}

Let us return to the controlled two-qubit gates \Eq{HctrlD}. It is convenient
for further applications to place the control qubit between the two 
controlled ones and then, rewriting \Eq{HctrlD}, it is possible to
consider a diagonal Hamiltonian with three energy levels
\begin{equation}
\begin{array}{rclrc}
 E&=&0 &:&
\ket{0}\ket{0}\ket{0},
\ket{0}\ket{0}\ket{1},
\ket{1}\ket{0}\ket{0},
\ket{1}\ket{0}\ket{1} \\
E&=&-\Delta E &:&
\ket{1}\ket{1}\ket{0},
\ket{0}\ket{1}\ket{1} \\
E&=&+\Delta E &:&
\ket{0}\ket{1}\ket{0},
\ket{1}\ket{1}\ket{1} 
\end{array}.
\label{EctrlD}
\end{equation}
The scheme has visual interpretation: if the control qubit is in the state $\ket{0}$,
the two data qubits are not interact, but if the control qubit
has the state $\ket{1}$, then the energy of the system is bigger for the data qubits 
in same states and smaller for different ones. So the program qubit may be considered
as a ``control switch'' of the interaction between the two data qubits. 
An illustrative, but a rather na{\"\i}ve scheme with a double quantum dot 
for the control of two spin systems is depicted on \Fig{qudcd}. 

Some discussion on physical systems appropriate for such a 
``switching'' purposes may be found also in papers related with 
models of the ``global'' quantum computing (with ``always-on interactions'')
\cite{Benj99,BenBos3,Benj4,BLR4}, because such approach is very close to 
the idea of programmable quantum networks.

\begin{figure}[htb]
\begin{center}
\IorII{\includegraphics[scale=0.4]{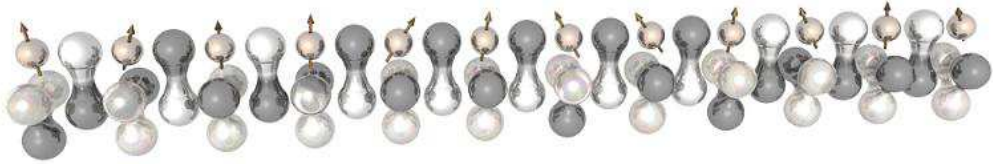}}
{\includegraphics[bb=0 0 960 160,scale=0.4]{quctr3x2n.jpg}}
\end{center}
\caption{A visual scheme of a programmable quantum register}
\label{Fig:qureg3x2}
\end{figure}

Finally, a programmable network with such a kind of gates for one- and 
two-qubit operations is depicted on \Fig{qureg3x2}. Here a data
register is presented as an array of spin-half systems and a control 
register is consisting of double and triple quantum dots. Double
quantum dots used for two-qubits operation here are intermittent with
data qubits, and qutrits are situated above.

The controlled one-qubit quantum gate for such a register may be described 
by Hamiltonians \Eq{HctrlXZ} and \Eq{Hctrit}
\begin{equation}
\op{h_{\rhd}^{\sf c}} = 
\ket{0}\bra{0}\otimes\op 1 + \ket{1}\bra{1}\otimes\op\sigma_x
+ \ket{2}\bra{2}\otimes\op\sigma_z.
\tag{\ref{Hctrit}$'$}
\label{Hctrit'}
\end{equation}
It is enough simply to include number $k$ of controlled qubit for
such Hamiltonian $\op{h_{\rhd}^{\sf c}}^{(k)}$. On the considered scheme 
\Fig{qureg3x2} for data qubits the number is $k=2j+1$.

Hamiltonians of controlled two-gates devote a bit more detailed
consideration, because one data qubit is controlled by
two different gates. Let us rewrite \Eq{decHd} for the control
qubits between the two data qubits
\begin{equation}
\op{h_{\between}^{\sf c}} = 
  \op\sigma_z \otimes \mat{0&0\\0&1} \otimes \op\sigma_z 
 = \frac{1}{2}\bigl(\op\sigma_z^{(1)}\op\sigma_z^{(3)} 
               -\op\sigma_z^{(1)}\op\sigma_z^{(2)}\op\sigma_z^{(3)}\bigr). 
\label{dcdH}
\end{equation}
The analogue of decomposition \Eq{decUd} is
\begin{multline}
\op{U_{\between}^{\sf c}}(\tau)=
\exp(i\op{h_{\between}^{\sf c}}\tau) =
  \exp(i\frac{\tau}{2}\op\sigma_z^{(1)}\op\sigma_z^{(3)})
 \exp(-i\frac{\tau}{2}\op\sigma_z^{(1)}\op\sigma_z^{(2)}\op\sigma_z^{(3)})\\
~ = \exp(i\frac{\tau}{2}\op\sigma_z^{(1)}\op\sigma_z^{(3)})
\, \exp(i\frac{\pi}{4}\op\sigma_z^{(1)}\op\sigma_x^{(2)})
\, \exp(-i\frac{\tau}{2}\op\sigma_y^{(2)}\op\sigma_z^{(3)})
\, \exp(-i\frac{\pi}{4}\op\sigma_z^{(1)}\op\sigma_x^{(2)}).
\label{dcdU}
\end{multline}

For an array with qubits Hamiltonians \Eq{dcdH} may be written as
\begin{equation}
\op{h_{\between}^{\sf c}}^{(2k+1)}   
 = \frac{1}{2}\bigl(\op\sigma_z^{(2k)}\op\sigma_z^{(2k+2)} 
               -\op\sigma_z^{(2k)}\op\sigma_z^{(2k+1)}\op\sigma_z^{(2k+2)}\bigr). 
\label{dcdHk}
\end{equation}
The Hamiltonians \Eq{dcdHk} commute for different $k$ and so despite
of the overlap, again there is no an essential difference between 
the control of array \Fig{qureg3x2} and the initial model with only three systems.

\subsubsection{Shift gates}

\begin{figure}[htb]
\begin{center}
\IorII{\includegraphics[scale=0.6]{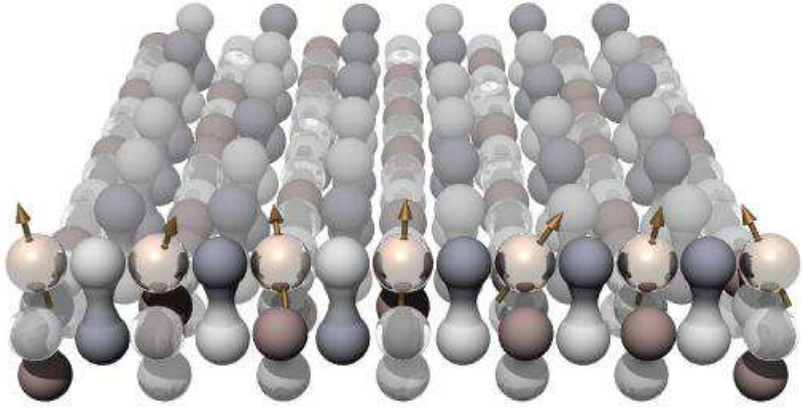}}
{\includegraphics[bb=0 0 512 256,scale=0.4]{quctr2D.jpg}}
\end{center}
\caption{A two-dimensional programmable quantum network}
\label{Fig:quctr2D}
\end{figure}

The models above represent only {\sf C} (control) part of the
{\sf Control-Shift} network discussed earlier in \Sec{3bus}. 
It is necessary to supply new and new indexes $\ket{k}$ for the control
register and if control and data buses may be considered as some 
arrays along a given axis ($x$) \Fig{qureg3x2}, it is possible to arrange 
a program bus as some bar along an orthogonal axis ($y$) with a separate 
line (1D array) for each control gate and to produce the two-dimensional 
array \Fig{quctr2D}.

In such 2D structure, each control element is receiving indexes 
from only one line, {\em i.e.}, 1D array along orthogonal axis 
($y$, program). Let us consider necessary operations with the array. 

The simplest idea --- is to implement the cyclic shift used \Eq{Shft} for
each such $y$-array
\begin{equation}
 \mathsf S \colon  \ket{k_L}\ket{k_{L-1}}\cdots\ket{k_2}\ket{k_1} 
           \mapsto \ket{k_1}\ket{k_L}\ket{k_{L-1}}\cdots\ket{k_2}.
\tag{\ref{Shft}$'$}
\label{yShft}
\end{equation}
Such a method produces valid realization of a programmable
{\sf Control-Shift} network, but there is a difficulty with
implementation of the Hamiltonian for \Eq{yShft}, because it
is {\em not a local} operation, {\em i.e.}, it acts on all $L$ systems.

On the other hand all Hamiltonians for the control above were
local with three systems or less. 
It is also possible to write local analogues of the {\sf Shift} operation.
The simple way --- is to consider two different operations: for one
step are exchanged all pairs $\ket{k_{2j+1}}$ and  $\ket{k_{2j+2}}$
and for next step --- pairs $\ket{k_{2j+2}}$ and $\ket{k_{2j+3}}$.

\begin{equation}
\setlength{\arraycolsep}{0pt}
\begin{array}{rrlcrl}
 \mathsf S_1 \colon& 
 \cdots\ket{k_{2j+1}}&\ket{k_{2j+2}}\cdots 
           &\ \mapsto\ &
 \cdots\ket{k_{2j+2}}&\ket{k_{2j+1}}\cdots 
\\
 \mathsf S_2 \colon&  
 \cdots&\ket{k_{2j+2}}\ket{k_{2j+3}}\cdots 
           &\ \mapsto\ & 
 \cdots&\ket{k_{2j+3}}\ket{k_{2j+2}}\cdots 
\end{array}
\label{SwpShft}
\end{equation}

So each step is performed using only the exchange ({\sf SWAP}) operation with 
two systems. For qubits, qutrits, and quantum systems with an arbitrary number
of states (``qu$d$its'') the {\sf SWAP} gate is defined on basis states as
\begin{equation}
{\sf SWAP} \colon \ket{k}\ket{l} \mapsto \ket{l}\ket{k}.
\label{swap}
\end{equation}
For qubits it may be also implemented using three 
{\sf c-NOT} gates \cite{FeyComp}
\begin{equation}
{\sf SWAP} =  \Mat{1&0&0&0\\0&0&1&0\\0&1&0&0\\0&0&0&1} =
\cNOT{12}\,\cNOT{21}\,\cNOT{12}.
\label{swap2}
\end{equation}

If to use the alternating sequence of such operations, states
with odd numbers like $\ket{k_1}$ are permanently shifted in one
direction and states with even numbers --- in opposite one. So it
is possible to encode a necessary sequence if to use only the odd states
and to set other to zero
\begin{equation}
 \ket{k_1}\ket{0}\ket{k_2}\ket{0} \cdots 
\label{k0k}
\end{equation}
Here is suggested as usual, that the zero index corresponds to the identity
operator. 

\subsection{Quantum cellular automata}
\label{Sec:QCA}

The present {\paper} is not devoted exclusively to the theory and applications of 
quantum cellular automata (QCA), but it is reasonable to discuss 
briefly the topic, because QCA models are widely used in many works
devoted to related problems. It was already briefly mentioned {\em the global
quantum computing} \cite{Benj99,BenBos3,Benj4,BLR4}. Recent works on 
reversible quantum cellular automata \cite{SchWer,VlQCA,Raus} make
possible to talk about even more direct relation between the model
of universal programmable quantum networks with pure states and 
quantum cellular automata.

The just considered model \Fig{quctr2D} illustrates, that the {\sf Control-Shift}
design has very close relation with cellular automata, but there
is specific subtleties for transition to the quantum case \cite{SchWer,VlQCA}. 

\begin{figure}[htb]
\begin{center}
\IorII{\includegraphics[scale=0.6]{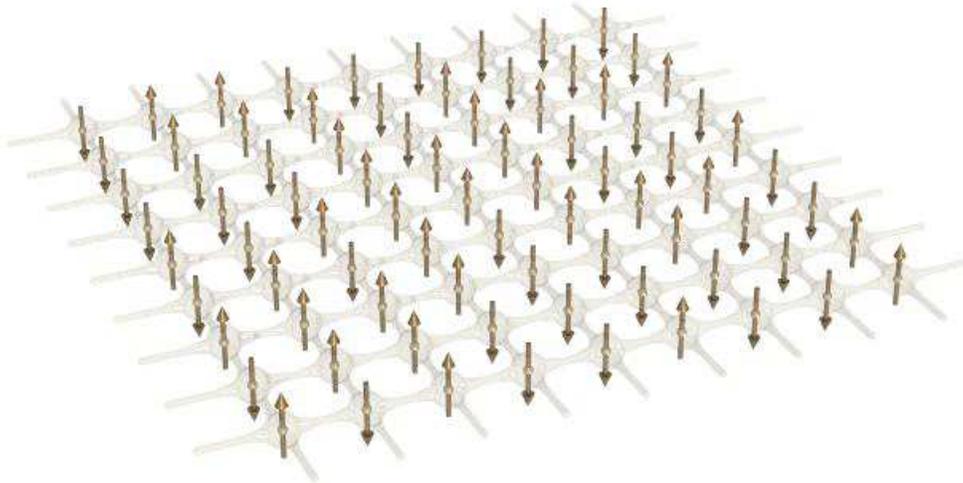}}
{\includegraphics[bb=0 0 640 320,scale=0.4]{qlattice.jpg}}
\end{center}
\caption{A lattice with qubits representing a cellular automaton}
\label{Fig:qlatt01}
\end{figure}

Let us consider for example a spin lattice. It could be used
formally for representation of a cellular automaton
with two values $\{0,1\}$ encoded by two basic states 
$\ket{\uparrow}$, $\ket{\downarrow}$ \Fig{qlatt01}. 

\begin{figure}[htb]
\begin{center}
\IorII{\includegraphics[scale=0.6]{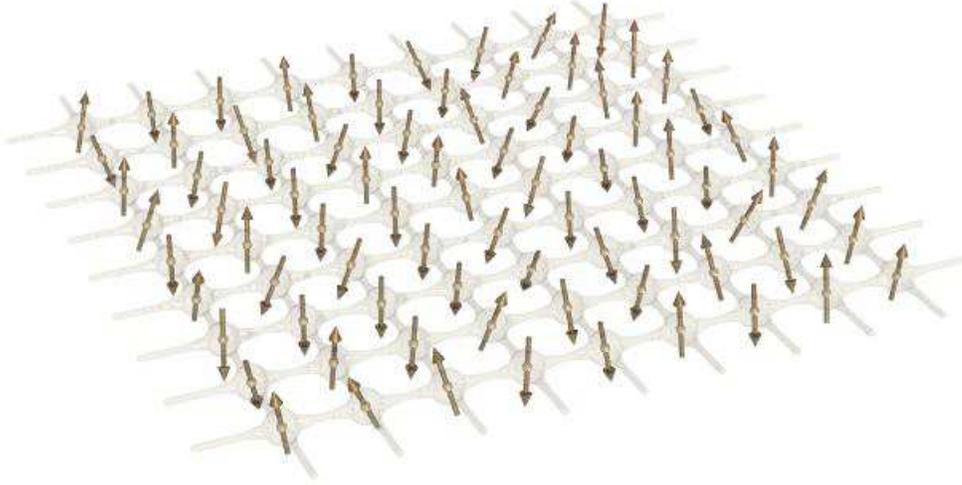}}
{\includegraphics[bb=0 0 640 320,scale=0.4]{qlattice2.jpg}}
\end{center}
\caption{A lattice with qubits described by the product state}
\label{Fig:qlattice}
\end{figure}

\begin{figure}[htb]
\begin{center}
\IorII{\includegraphics[scale=0.3]{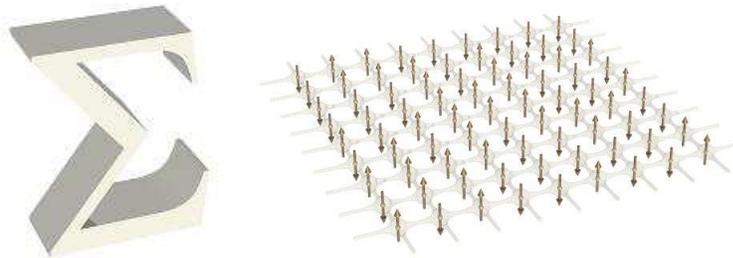}~
\includegraphics[scale=0.3]{qlattice.eps}}
{\includegraphics[bb=0 0 280 320,scale=0.2]{sumlat.jpg}~
\includegraphics[bb=0 0 640 320,scale=0.2]{qlattice.jpg}}
\end{center}
\caption{A sum on different configurations representing
 general entangled state}
\label{Fig:qlatsum}
\end{figure}

On the other hand, a general state of such a lattice is not necessary
may be represented as a {\em product state} \Fig{qlattice}, then
it may be simply treated as a collection of cells in different states.
The general entangled state is defined as a sum \Fig{qlatsum} with complex
coefficients on different basis configurations \Fig{qlatt01}. 

This specific property of a quantum system does not permit to talk about
a state of a single cell, or a block of cells as about a vector in the Hilbert
space, and it produces specific difficulties for local description 
of cellular automata. 

On the other hand, it was already mentioned, that the program bus may be 
formally described using {\em reversible} classical computations and 
the theory of reversible cellular automata has a long history and
is appropriate for such a purpose \cite{TT77,TM90,WCA}.

One specific method used in the theory of reversible cellular automata
\cite{TM90} is the {\em Margolus partition} with two different operations
during different time steps. The simplest example for 1D automata
just corresponds to \Eq{SwpShft} used for local implementation
of the shift operation. 

The alternating sequence of {\sf S} and {\sf C} operations in \Eq{cyCS} 
used above in programmable {\sf Control-Shift} networks in \Sec{3bus} 
also becomes a quantum analogue of the Margolus partition, if to consider 
2D design of quantum gates described below \Fig{quctr2D} as a quantum
cellular automaton. 

The design considered here maybe still does not look like ``true'' QCA, 
because it is not regular enough, {\em e.g.}, there are 2D array
for program with two different kinds of systems and 1D array of data.
On the other hand, cellular automaton with different kinds of cells
is equivalent to cellular automaton with only one kind with additional flag. 

\begin{figure}[htb]
\begin{center}
\IorII{\includegraphics[scale=0.6]{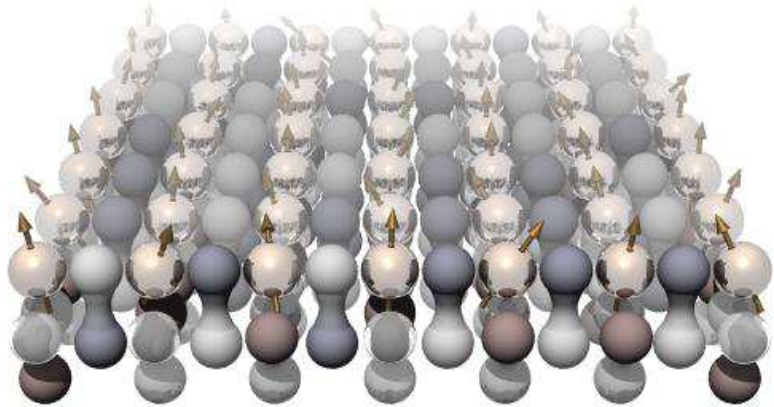}}
{\includegraphics[bb=0 0 512 256,scale=0.4]{quctr2D2.jpg}}
\end{center}
\caption{A quantum network with two-dimensional data and program
registers}
\label{Fig:quctr2D2}
\end{figure}

It is also useful to consider 2D array of data \cite{Raus}. We
just need to spread the whole programmable register \Fig{qureg3x2} 
along program bus \Fig{quctr2D2} (instead of spreading only control 
elements of the register depicted earlier on \Fig{quctr2D}).

Such array may be considered as many independent copies of the 
programmable register. To produce initial design \Fig{quctr2D} from 
such a regular QCA, it is possible formally to introduce an
additional flag-qubit and to modify the control register in such a way, 
that it changes state of data only if the flag is $\ket{1}$.

\begin{figure}[htb]
\begin{center}
\IorII{\includegraphics[scale=0.6]{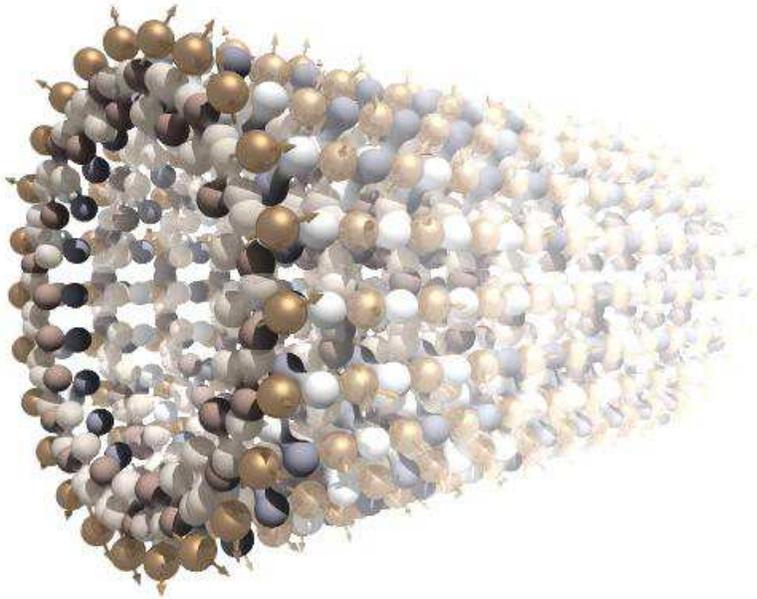}}
{\includegraphics[bb=0 0 512 384,scale=0.4]{quctr2DC.jpg}}
\end{center}
\caption{A programmable quantum network as cylindrical QCA}
\label{Fig:quctr2DC}
\end{figure}

On the other hand, it is possible to save the initial structure and
to let all program registers evolve in parallel. Implementation
of the cyclic shift may be represented by cylindrical QCA \Fig{quctr2DC}.

Here we have three basic elements: the qubits of quantum data
register(s), triple quantum dots for implementation of one-qubit
gates (below a qubit on \Fig{qureg3x2}, \Fig{quctr2D2}, {\em etc.}), 
and double quantum dots for control of two-qubit gates (between each 
two qubits on \Fig{qureg3x2}, it should be mentioned, that
on \Fig{quctr2D2} the registers are arranged {\em along the axis} of 
the cylinder).

The QCA programmable network uses two standard steps {\sf C} and {\sf S}
of {\sf Control-Shift} network, but they are subdivided on smaller steps.
The analysis of operation {\sf S} is simpler due to direct relation with
reversible classical computations already mentioned above.

The {\sf S(hift)} steps move data around cylinder and may be implemented
using two sub-steps $\mathsf{S_1, S_2}$ \Eq{SwpShft}. It was
already mentioned, that the indexes in such a case should be
intermittent with zeros \Eq{k0k}.

Each control line may be considered independently and so {\sf S}
step is described by the expression \Eq{SwpShft}. For implementation
of the universal set of gates like \Eq{gZXD} two different kinds of 
lines are necessary: for the control of two-qubit operations
are used only two indexes $\{0,1\}$ and for controlled one-qubit gates 
(at least for two first qubits) are used three indexes
$\{0,1,2\}$. 

The lines with indexes for one-qubit gates are alternating with
lines for two-qubit gates. It is important to mention, that all
two-qubit gates from \Eq{gZZk} are commuting and so may be applied 
on the same step. The one-qubit gates on different lines are also
commuting. On the other hand, gates \Eq{gXk} are not commuting with 
gates \Eq{gZZk} for adjoint lines, and so it is necessary to 
apply such gates on different steps. It requires a special
arrangement of nonzero indexes on adjoint lines.

Let us suggest that such arrangement condition is satisfied 
to prevent consideration of non-commuting Hamiltonians and consider 
{\sf C(ontrol)} step. 

One-qubit operations may be defined locally 
for each pair with the control qutrit and the data qubit. The Hamiltonian 
of control on such a pair was already discussed above \Eq{Hctrit'}
on the page \pageref{Hctrit'}.
The Hamiltonians of controlled two-qubit gates again coincide 
with expressions for single controlled register \Fig{qureg3x2} 
and also were considered earlier \Eq{dcdHk}. 

It was already mentioned, that the Hamiltonians \Eq{dcdHk}
for control commute for different sites. The commutativity
is important for quantum cellular automata models \cite{SchWer,VlQCA}. 
Let us suggest, that an entire Hamiltonian of QCA may be
expressed as a sum of all local Hamiltonians corresponding to
local cells. Only for commutative operators such a method produces
appropriate relation between global and local transition
functions
\begin{equation}
 \exp(i\sum_{k,j}\op h^{k,j} \Delta\tau)
 =\prod_{k,j}\exp(i\op h^{k,j}\Delta\tau), \quad
 \op U_{\text{global}} = \prod_{k,j}\op U^{k,j}_{\text{local}}
\label{UprodQCA}
\end{equation}

The Margolus partition used in the theory of reversible QCA \cite{SchWer}
just uses two sets of commuting Hamiltonians. In the model considered
here may be found even four such sets. The Margolus partition
used for {\sf S(hift)} operation is corresponding to the two operators 
$\mathsf{S_1, S_2}$ \Eq{SwpShft} and already was discussed above,
but {\sf C(ontrol)} operation also contains two sets of
Hamiltonians corresponding to one- and two-gates
\begin{equation}
\op{H_{\rhd}^{\sf c}} = \sum_{k,j}\op{h_{\rhd}^{\sf c}}^{(2k-1,j)}, \quad
\op{H_{\between}^{\sf c}} = \sum_{k,j}\op{h_{\between}^{\sf c}}^{(2k,j)}.
\label{SumH2D}
\end{equation}
Let us introduce two operators
\begin{equation}
{\mathsf C}_1 = \exp(i \op{H_{\rhd}^{\sf c}} \Delta\tau),
\qquad
{\mathsf C}_2 = \exp(i \op{H_{\between}^{\sf c}} \Delta\tau),
\label{C1C2}
\end{equation}
with Hamiltonians of one- and two-gates respectively and $\Delta\tau$
is the fixed irrational multiple of $\pi$, then the programmable quantum
network based on such QCA may be described by the periodic sequence of
operators like
\begin{equation}
\mathsf{U_{VI}} = \mathsf{ C_1 C_2 S_1 \: C_1 C_2 S_2}, \qquad
\mathsf{U_{IV}} = \mathsf{C_1 S_1 \: C_2 S_2},
\label{U6U4}
\end{equation}
there the second operator $\mathsf{U_{IV}}$ with only four terms is 
using alternating (`checker-board') arrangement, when after each step 
half of indexes are zeros and so only one operator between $\mathsf{C_1, C_2}$ 
produces nontrivial result. It should be mentioned, that both
operators \Eq{U6U4} correspond to two steps of a program and
so for the cyclical design with the perimeter $2L$, it is necessary
repeat such operators $L$ times.

\section{Conclusion}

In this {\paper} was considered the theory of programmable quantum
networks. They may be considered as quite appropriate models of 
future quantum processors. Such devices would not require 
the macroscopic equipment traditional for modern experimental
research in the area of quantum information processing,
because it is suggested to use quantum systems for the
program, control and data. 

Such architecture is also interesting from the pure theoretical
point of view. It let us use the unified description of all 
processes necessary for functionality of such quantum 
devices and reduces amount of problems related with consideration
of transition from classical to quantum domain. 

It should be mentioned, that some methods used in this {\paper}
also may be found in the more general theory about relations
between quantum and classical pictures \cite{Zur98,Zur03} and so application
for the quantum information science is quite justified.

In the present {\paper} were used programmable quantum networks 
with pure states (sometime also called `deterministic') and 
it produces additional clarification and simplification. 
It is also actual because `probabilistic' programmable quantum 
gate arrays already are presented quite completely in other works.

The theory of programmable quantum networks is still in the state 
of development and some new interesting branches may appear. 
For example universal programmable quantum computers by definition 
may perform any transformations with data, but it corresponds to arbitrary
manipulations with a quantum system. Due to such a principle
emphasized already in earliest works \cite{UQC} there is no big 
difference between an universal programmable quantum network and 
a quantum robot \cite{QR98} with wide range of possible operations. 

Say, it is quite reasonable to consider
the famous Wigner's question about possibility of ``self-reproducing 
units'' using the theory of universal programmable networks. 
In fact, some preliminary analysis was already performed using 
both probabilistic \cite{QmUc} and deterministic \cite{NRQU}
approach, but the intriguing problem most likely devotes further research.

\section*{Acknowledgments}
Author is grateful to V. Bu\v{z}ek, R. Raussendorf and A. K. Pati 
for the interesting exchange.

\end{document}